\documentclass[aps,prx,twocolumn,superscriptaddress,floats,floatfix]{revtex4-1}
\usepackage{amssymb,amsmath,amsthm,bm}
\usepackage{graphicx}
\usepackage{epstopdf}
\usepackage{color}
\usepackage{subfigure}
\usepackage{epsfig}
\usepackage{rotating}
\usepackage{mwe}
\usepackage{float}
\usepackage{mathrsfs,multirow}
\usepackage{dcolumn}
\usepackage{verbatim}
\usepackage{hyperref}
\usepackage[Bistream charter]{mathdesign}

\usepackage[T1]{fontenc}
\usepackage[latin9]{inputenc}
\usepackage[a4paper]{geometry}
\usepackage{array}
\usepackage{multirow}
\usepackage{esint}
\usepackage{babel}

\def \Sp {S_{\rm p}}

\def \Fp {F_{\rm p}}
\def \FQLp {F^{\rm QL}_{\rm p}}
\def \FEup {F^{\rm Eu}_{\rm p}}
\def \SQLp {S^{\rm QL}_{\rm p}}
\def \SEup {S^{\rm Eu}_{\rm p}}
\def \Tint {T^{\rm I}_{\rm p}}
\def \TintQL {T^{\rm I,\,QL}_{\rm p}}
\def \TintEu {T^{\rm I,\,Eu}_{\rm p}}
\def \chiInt {\chi^{\rm I}_{p}}
\def \chiQL {\chi^{\rm I,\,QL}_{p}}
\def \chiEu {\chi^{\rm I,\,Eu}_{p}}
\def \zetaQL {\zeta^{\rm QL}}
\def \zetaEu {\zeta^{\rm Eu}}
\def \taud {\tau_{\rm up}}
\def \tfm {\tau_{\rm col}}

\def \Tpfm {T^{\rm p}_{\rm col}}
\def \TpfmA {T^{\rm p}_{{\rm col},A}}
\def \TpfmBI {T^{\rm p}_{{\rm col},B1}}
\def \TpfmBII {T^{\rm p}_{{\rm col},B2}}

\def \zpfm {z^p_{\rm col}}
\def \Li {L_{\rm I}}
\def \kmax {k_{\rm max}}
\def \urms {u_{\rm rms}}

\def \tstar {t_{\ast}}

\newcommand{\bra}[1]{\left\langle #1\right\rangle}
\def \TL {T_{\rm L}}
\newcommand{\eq}[1]{~(\ref{#1})}
\begin{document}

\title{Dynamic multiscaling in stochastically forced Burgers turbulence}

\author{Sadhitro De}
\email{sadhitrode@iisc.ac.in}
\affiliation {\footnotesize Centre for Condensed Matter Theory, Department of Physics, Indian Institute of Science, Bangalore 560012, India}

\author{Dhrubaditya Mitra}
\email{dhruba.mitra@gmail.com}
\affiliation{\footnotesize NORDITA, KTH Royal Institute of Technology and Stockholm University, 
Roslagstullsbacken 23, 10691 Stockholm, Sweden}

\author{Rahul Pandit}
\email{rahul@iisc.ac.in}
\affiliation{\footnotesize Centre for Condensed Matter Theory, Department of Physics, Indian Institute of Science, Bangalore 560012, India}

\begin{abstract}
We carry out a detailed study of dynamic multiscaling in the turbulent nonequilibrium, but statistically steady, state of the stochastically forced one-dimensional Burgers equation. 
We introduce the concept of \textit{interval collapse times} $\tfm$, the time taken for an interval of length $\ell$, demarcated by
a pair of Lagrangian tracers, to collapse at a shock. By calculating
the dynamic scaling exponent of the order-$p$ moment of $\tfm$,
we show that (a) there is \textit{not one but an infinity  of characteristic
time scales} and (b) the probability distribution function of $\tfm$ is non-Gaussian and has a power-law tail. Our study is based on (a) a theoretical framework that allows us to obtain dynamic-multiscaling exponents analytically, (b) extensive direct numerical simulations, and (c) a careful comparison of the results of (a) and (b). We discuss possible generalizations of our work to dimensions $d >1 $, for the stochastically forced Burgers equation, and to other compressible flows that exhibit turbulence with shocks.
\end{abstract}

\maketitle

\section{Introduction}\label{sec:Intro}

Studies of statistically homogeneous and isotropic turbulence, one of the most important examples of a non-equilibrium steady state (NESS), lie at the interfaces between non-equilibrium statistical mechanics, fluid dynamics, and spatiotemporal chaos. In this NESS, the turbulent fluctuations of the flow velocity span wide ranges of spatial and temporal scales; and  the correlation or structure functions, which characterize these fluctuations, display power laws in space and time~\cite{frisch,HayotDynStr,HayotPhysFlu,SpatStr1,MitraDyn,SSR_NJP,SSR_PRL,PanditDynRev}. It has been suggested several times that these power-law forms are like those seen in correlation functions in equilibrium critical phenomena~\cite{chaikin,kardar2007statistical,goldenfeld2018lectures,forster1977large,dedominicis1979energy}. Perhaps the simplest example of a power-law form in such turbulence is the scaling of the energy spectrum, $E(k)\sim k^{-\alpha}$, for wave numbers $k$ in the \textit{inertial range} $\eta^{-1}\ll k \ll\Li^{-1}$,
where $\eta$ and $\Li$ are, respectively, the dissipation and energy-injection length scales~\cite{frisch,pandit2009statistical}.
The phenomenological theory, proposed by Kolmogogorov in 1941 (K41), yields the exponent $\alpha^{K41} = 5/3$, but the intermittency of turbulence leads to multifractal corrections to this and to other exponents (see below and, e.g.,
Refs.~\cite{frisch,pandit2009statistical,parisi1985multifractal,benzi1984multifractal}).  To understand this multifractality we must go well beyond~\cite{frisch,pandit2009statistical,parisi1985multifractal,benzi1984multifractal} the theoretical framework that is used to explain power-law correlation functions at equilibrium critical points~\cite{chaikin,kardar2007statistical,goldenfeld2018lectures}.

Multifractal fluctuations lead to considerable theoretical challenges when we move from equal-time to time-dependent correlation functions~\cite{HayotDynStr,HayotPhysFlu,SpatStr1,MitraDyn,SSR_NJP,SSR_PRL,PanditDynRev}. In simple critical phenomena,
power-law forms for time-dependent correlation functions come from the divergence of the correlation time $\tau$ that is related to the diverging correlation length $\xi$ by the \textit{dynamic scaling Ansatz}~\cite{halperin1967generalization,ferrell1968fluctuations,hohenberg1977theory}, $\tau\sim \xi^\theta$, with $\theta$ the \textit{dynamic scaling exponent}. Multifractal velocity fluctuations in turbulence lead to dynamic multiscaling of time-dependent correlation functions for which we must use an infinity of \textit{dynamic scaling Ans\"atze}. This has been discussed in detail in a variety of hydrodynamical partial differential equations (PDEs)~\cite{frisch,HayotDynStr,HayotPhysFlu,SpatStr1,MitraDyn,SSR_NJP,SSR_PRL,PanditDynRev,MHDShellDyn}.

We elucidate dynamic multiscaling in the NESS of the stochastically forced Burgers equation. We show that the theoretical methods that have been used to study such multiscaling of turbulence in incompressible hydrodynamical PDEs \textit{cannot}
be used for turbulence in the stochastically forced Burgers equation (often referred to as Burgers turbulence)~\cite{frisch2001burgulence,Bec2007BurgersT,HayotStrFn,HayotMultifrac,BurgMitra}, which is \textit{compressible}. Therefore, we introduce an infinity of time scales that can be used to characterize the spatiotemporal evolution of Burgers turbulence in $d$ spatial dimensions. Our study is based on (a) a theoretical framework, for $d=1$, that allows us to obtain dynamic multiscaling exponents analytically and (b) extensive direct numerical simulations (DNSs) for $d=1$ and $d=2$. 
To the best of our knowledge, \textit{this is the first time that such multiscaling exponents have been obtained analytically for turbulence in a nonlinear hydrodynamical PDE.} Before we present the details of our study, we give a qualitative overview of our work and its significance in the light of earlier studies of related problems.

We can extract time scales from turbulent flows in a variety of different ways. The K41 theory~\cite{K41} suggests simple dynamic scaling, i.e., all time scales are characterized by the same dynamic exponent $\theta^{K41} = 2/3$. However, from DNSs and a heuristic understanding of various models of turbulence, we have learned that incompressible turbulence exhibits \textit{dynamic multiscaling}, i.e., different time scales are linked to length scales via different dynamic exponents~\cite{SpatStr1,BiferaleCorr,MHDShellDyn,MitraDyn,SSR_NJP,SSR_PRL,PanditDynRev}.  L'vov, Podivilov, and Procaccia~\cite{SpatStr1} had proposed that the characteristic times of eddy-velocity correlation functions of various orders are linked to the eddy sizes through multiple dynamic exponents, which can be related to the equal-time structure function exponents by linear bridge relations. 
Thereafter, Mitra and Pandit~\cite{MitraDyn} not only confirmed this from their shell-model DNS, but also showed that the dynamic exponents depend specifically on how the time scales are defined. In addition, for the passive-scalar problem, Mitra and Pandit~\cite{mitra2005dynamics} showed that simple dynamic scaling is obtained if the velocity field is of the type in the Kraichnan model~\cite{Kraichnan,falkovich2001particles}; but \textit{bona fide} dynamic multiscaling, with an infinity of nontrivial dynamic exponents, is obtained only if the advecting velocity itself exhibits dynamic multiscaling. We note that the analysis of such passive-scalar turbulence is much simpler than its fluid-turbulence counterpart because the passive-scalar equation is \textit{linear} in the scalar concentration. For fluid turbulence, we must confront the nonlinearity of the incompressible Navier-Stokes equations.

Typical shell models, which are coupled ordinary differential equations (ODEs) designed to mimic turbulence in hydrodynamical PDEs, are defined in a logarithmically discretised wave-number (Fourier) space, with shells, labelled by the wave number $k$,
and the complex, scalar shell velocity $u_k$. Given their  simplicity, such shell models cannot yield reliable flow fields; however, they have been remarkably successful in obtaining turbulent states with spectral and multifractal properties akin to those that are obtained for turbulence in their parent hydrodynamical PDEs (see, e.g., Refs.~\cite{pandit2009statistical,biferale2003shell}). The interaction between velocities in shell models is in marked contrast to the interaction of Fourier modes of the velocities in hydrodynamical PDEs. For instance, in the incompressible Navier-Stokes equation, the nonlinear term couples all velocity Fourier modes to each other, whereas shell-model velocities $u_k$ interact only with velocities in nearest- or next-neighbor shells, so small-$k$ velocities do not affect large-$k$ velocities \textit{directly}. By contrast, in hydrodynamical PDEs, small eddies (large-$k$ modes) are advected directly by large ones (small-$k$ modes), because of the coupling of all Fourier modes to each other, whence we get a \textit{sweeping effect}~\cite{Sweep} that yields eddy lifetimes that are linearly proportional to the eddy size, i.e., the dynamic exponent is unity. This sweeping effect, which masks the underlying dynamic multiscaling of turbulent flows, is a manifestation of Taylor's hypothesis.

For incompressible turbulence, Belinicher and L'vov~\cite{QL1} had suggested that quasi-Lagrangian (QL) velocities, calculated in the reference frame of a Lagrangian particle or tracer, should be free from sweeping effects. This was shown explicitly by Ray, Mitra, Perlekar, and Pandit~\cite{SSR_PRL}, who quantified the dynamic multiscaling of forward-cascade turbulence in the incompressible two-dimensional ($d=2$) Navier-Stokes (NS) equation. Biferale, Calzavarini, and Toschi~\cite{BiferaleCorr} used QL velocities to obtain dynamic multiscaling for turbulence in the incompressible three-dimensional ($d=3$) Navier-Stokes equation. Ray, Mitra, Perlekar, and Pandit~\cite{SSR_PRL} also showed that sweeping effects can be suppressed by friction, which removes energy from a flow at all spatial scales.

The characterization of dynamic multiscaling in incompressible-fluid turbulence via the QL approach cannot be used when we consider turbulence in a compressible fluid. The Lagrangian particles, which we require to define QL fields, get trapped in shocks, which form in compressible turbulent flows. We show this explicitly by DNSs for turbulence in the one-dimensional ($d=1$) stochastically forced Burgers equation. Before we show how to overcome this difficulty, it is useful to recall some elementary results for Burgers turbulence.

The Burgers equation, the simplest compressible hydrodynamical PDE, which was introduced for the study of fluid equations in the low-viscosity limit and then for pressure-less gas dynamics~\cite{bateman1915some,burgers2013nonlinear}, has the same nonlinear term as the NS equation; so it is often used as a testing ground for statistical theories of turbulence~\cite{frisch2001burgulence,Bec2007BurgersT}. The $d$-dimensional Burgers equation can be solved via the
Hopf-Cole transformation~\cite{hopf1950partial,cole1951quasi,burgers2013nonlinear}; in the inviscid limit, this yields a maximum principle for a  velocity potential~\cite{gurbatov1997,frisch2001burgulence,Bec2007BurgersT}. In cosmology, the Burgers equation is used to model the formation and distribution of large-scale structures in the universe~\cite{Vergassola}, under the \textit{adhesion approximation~}\cite{Adhesion}. 

Studies of Burgers turbulence (sometimes referred to as Burgulence) comprise investigations of the statistical properties of (a) solutions of the Burgers equation with random initial data or (b) solutions of the Burgers equation with stochastic forcing~\cite{HayotScaling,Chekhlov,BurgMitra}; we concentrate on the latter. The stochastically forced Burgers equation can be derived by taking a spatial derivative of the Kardar-Parisi-Zhang (KPZ) equation~\cite{KPZmain,KPZ1}, which models the height profile of a growing interface. For the physically relevant forms of the stochastic forcing used in the KPZ equation, correlation functions display simple scaling of equal-time and time-dependent correlation functions.

We consider turbulence in the $d=1$ Burgers equation with a zero-mean, white-in-time, Gaussian random force whose variance, in Fourier space, scales as $\sim k^{-\beta}$. This stochastically forced equation yields a NESS with properties that are akin to those in $d=3$ NS turbulence. In particular, one-loop renormalization-group (RG) calculations yield a K41-type energy spectrum 
$E(k) \sim k^{-5/3}$ for $\beta = 1$, for $k$ in the inertial range~\cite{Chekhlov,PolyakovRG,HayotDynStr,HayotMultifrac,HayotPhysFlu,HayotScaling,HayotStrFn,BurgMitra}.
DNSs show that velocity structure functions (see below) in this NESS display multiscaling~\cite{Chekhlov,PolyakovRG,HayotDynStr,HayotMultifrac,HayotPhysFlu,HayotScaling,HayotStrFn,BurgMitra}. It has been conjectured that this multiscaling could be a numerical artifact~\cite{BurgMitra} which could wane with increasing spatial resolution of the DNS, and be replaced by bifractal scaling.

Given the challenges that we have outlined above, the study of dynamic scaling in the stochastically forced Burgers equation is not as well developed as it is in incompressible fluid turbulence~\cite{SpatStr1,BiferaleCorr,MHDShellDyn,MitraDyn,SSR_NJP,SSR_PRL,PanditDynRev}. An earlier DNS-based study of certain time-dependent, Eulerian velocity structure functions for this model yielded a single dynamic exponent of unity~\cite{HayotDynStr} that was attributed to the sweeping effect. We go beyond this Eulerian study and the 
QL investigations of incompressible fluid turbulence~\cite{SpatStr1,BiferaleCorr,MHDShellDyn,MitraDyn,SSR_NJP,SSR_PRL,PanditDynRev}.

As we have noted above, Lagrangian tracers get trapped at shocks in compressible turbulent flows, because of which the QL transformation might not be adequate for the removal of sweeping effects. For turbulence in the $d=1$ stochastically forced Burgers equation, we overcome this difficulty by using \textit{a pair of tracers} separated initially by a Lagrangian interval of length $\ell$. We then compute the \textit{interval-collapse time}, $\tfm$, which we define as the time at which this pair collapses to a point at a shock. We find that $\tfm$ depends on both $\ell$ and the location of the interval. Hence we compute, for each value of $\ell$, the probability distribution function (PDF) of $\tfm$ and extract a hierarchy of time scales, $\Tpfm(\ell)$, from its order-$p$ moments. We make the dynamic-scaling \textit{\text{Ans\"atze}}, $\Tpfm(\ell)\sim \ell^{\zpfm}$, and obtain therefrom the \textit{interval-collapse exponents} $\zpfm$ for different values of $p$. We find from our high-resolution DNS that $\zpfm$ is not a linear function of $p$. This indicates dynamic multiscaling and intermittency. We develop a theory for the $p$-dependence of $\zpfm$, which indicates that $\zpfm$ should be a piecewise linear function of $p$, i.e., we obtain a particular case of dynamic multiscaling, namely, dynamic bi-scaling. Our theory also yields an analytical expression for the PDF of $\tfm$. The PDF is non-Gaussian and has a power-law tail; we compare this with the results of our DNS. We also show from our DNS that the QL approach is indeed unable to suppress sweeping effects. Furthermore, we examine the extension of our \textit{interval-collapse} framework for examining dynamic scaling in the stochastically forced Burgers equation to dimensions $d > 1$ and the challenges in carrying out such a study.

The remainder of this paper is organized as follows: In Sec.~\ref{sec:Model} we define the stochastically forced Burgers equation and outline the numerical methods that we use. In Sec.~\ref{sec:colltimes}, we investigate the statistical properties of the interval-collapse times, $\tfm$, and show how to calculate the interval-collapse exponents $\zpfm$. In Sec.~\ref{sec:QLstr}, we explore dynamic multiscaling via QL velocity structure functions. Finally, in Sec.~\ref{sec:Discussions}, we discuss the significance of our results and propose ways of extending our approach to $d > 1$ and compressible turbulent flows with shocks. 

\section{Model and Numerical Methods}\label{sec:Model}
\begin{figure}[t]
    \centering
    \includegraphics[scale=0.63]{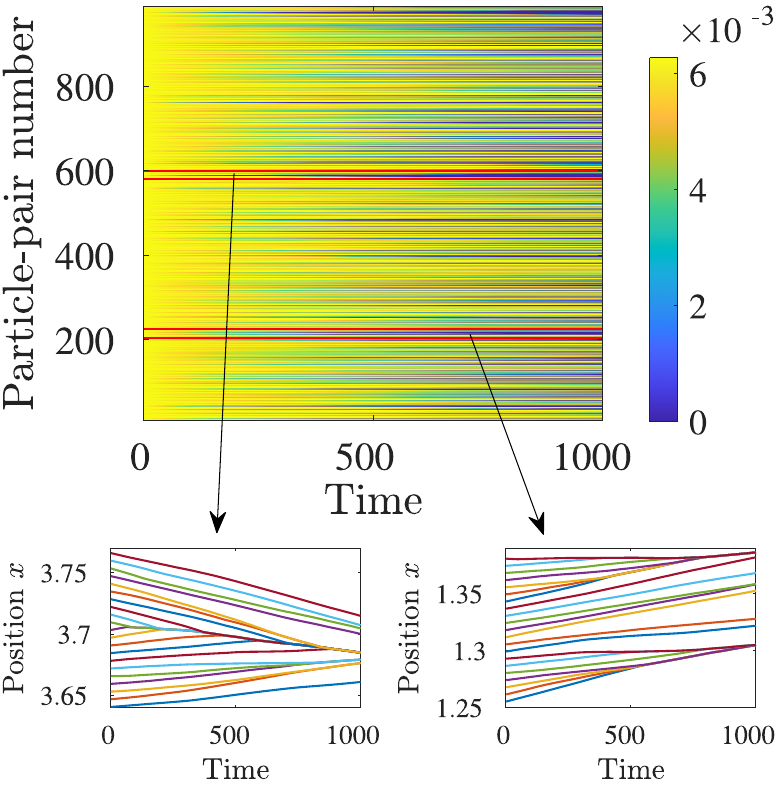}
    \caption{\small (Color online) Space-time plots of representative tracer trajectories (run $\rm{R1}$) for initially equi-spaced tracers; we choose $1000$ tracers for ease of visualization here. The vertical axis gives the nearest-neighbor tracer-pair number (the tracers are numbered in ascending order and the pair number $i$ denotes two tracers that are initially at $2\pi i/1000$ and $2\pi (i+1)/1000$); the horizontal axis denotes the time (in units of iteration number). The color bar denotes the separation between the tracers within each of the nearest-neighbor tracer pairs. As time progresses, the separation between the tracers in such pairs decreases. \textit{Bottom panel:} Expanded versions of the tracks of the tracers (distinguished by different colors) enclosed within the red-bordered rectangular boxes in the top panel.}
    \label{fig:mergeLag}
\end{figure}
The $d=1$ stochastically forced Burgers equation that we consider is
\begin{equation}
\partial_tu+u\partial_xu=\nu\partial_{xx}u+f(x,t)\;\;,
\label{eq:Burgers}
\end{equation}
where $u(x,t)$ is the fluid velocity at position $x$ and time $t$, $\nu$ is the kinematic viscosity, and $f$ is a zero-mean, Gaussian white-in-time random force whose Fourier components $\hat{f}(k,\,t)$ satisfy 
\begin{equation}
\langle \hat{f}(k,\,t)\hat{f}(k',\,t')\rangle\sim k^{-\beta}\delta(k+k')\delta(t-t'),
\label{eq:noise}
\end{equation}
where $k$ is the the wave number. We choose $\beta=1$, because, at the level of a one-loop RG, this choice of $\beta$ yields a K41-type energy spectrum $E(k) \equiv \langle |\hat{u}(k,t)|^2 \rangle \sim k^{-5/3}$, for $k$ in the inertial range, where $\hat{u}(k,t)$ is the Fourier transform of $u(x,t)$, and the angular brackets denote a time average over the NESS~\cite{HayotMultifrac,HayotScaling,HayotDynStr,BurgMitra}. Here, $\eta=(\nu^3/\epsilon)^{1/4}$ is the dissipation scale beyond which viscous losses are significant, $\epsilon$ is the mean energy dissipation rate, and $\Li \equiv \sum_k[E(k)/k]\big/\sum_kE(k)$ is the integral length scale.

In our DNSs of Eqs.~(\ref{eq:Burgers}) and
(\ref{eq:noise}), we use periodic boundary conditions in a domain of length $L=2\pi$ and $N$ collocation points. To achieve high spatial resolutions we use $N=2^{16}$ (Run R1) and $N=2^{20}$ (Run R2). We employ a standard pseudospectral method with the $2/3$-dealiasing rule~\cite{canuto2007spectral}. For time-stepping we use the implicit Euler-Maruyama scheme~\cite{kloeden1992stochastic,higham2001algorithmic}. We generate the stochastic force $\hat{f}(k,t)$ in Fourier space, with a high-wave-number cutoff at $k_c = N/8$. We have confirmed that our results remain unchanged even if we evolve the unforced equation by using the second-order exponential time-differencing Runge-Kutta method and then add the forcing term, $\hat{f}(k,t)\sqrt{\delta t}$, to the velocity field at the end of every time step of step size $\delta t$. The parameters for our DNS runs are given in Appendix~\ref{app:A}. Most of our DNSs and data analysis have been carried out on a GPU cluster with NVIDIA Tesla K20 accelerators. Once the system reaches its NESS, the energy spectrum $E(k)$ shows inertial-range scaling over two decades in $k$; a typical snapshot of a steady-state velocity profile and a compensated plot of $E(k)$, for run R2, are shown in the Appendix~\ref{app:A}.

After our system reaches its NESS, we introduce $N_p$ equi-spaced Lagrangian particles (or tracers), whose equations of motion are
\begin{equation}
    \frac{d}{dt}R_i(t)=U_i\quad{\rm and}\quad U_i=u(x,t)\delta(x-X_i) \, ,
    \label{eq:tracer}
\end{equation}
where $X_i$ and $U_i$ are the instantaneous position and velocity, respectively, of the $i-$th tracer, and $\delta(x)$ is the Dirac delta function. We solve (\ref{eq:tracer}) by using the forward-Euler method. The $\delta(x-X_i)$ factor is implemented by linear interpolation. As time progresses, the tracers cluster at the shocks, as we show in Fig. \ref{fig:mergeLag}.

\section{Interval-collapse Times}\label{sec:colltimes}
\begin{figure*}[t]
\begin{centering}
\includegraphics[scale=0.335]{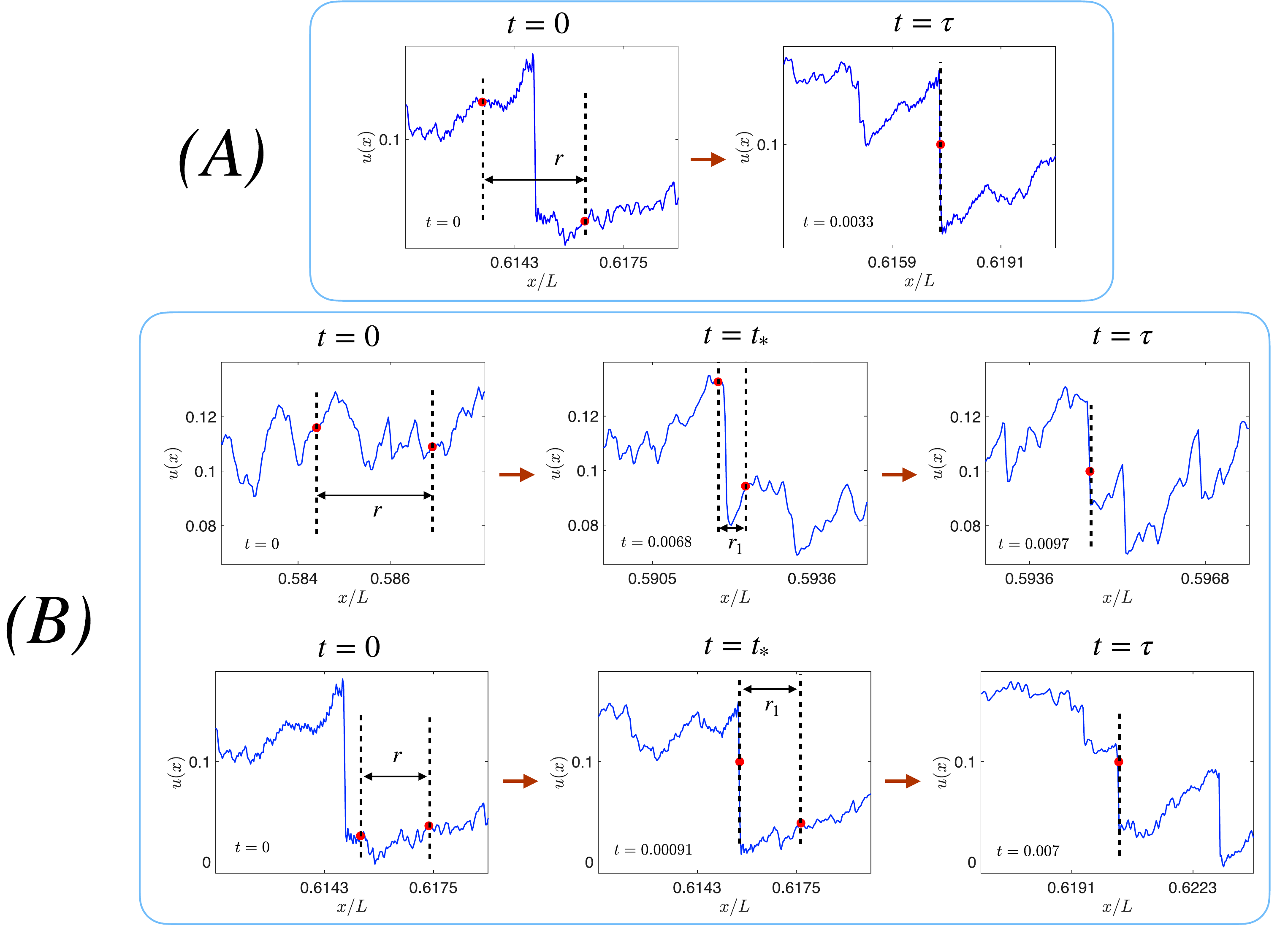}
\par\end{centering}
\caption{\label{fig:tfmillus}{\small Examples of the possible cases of collapse of an interval of initial length $r$ from our DNSs: \textit{Panel (A)} The collapsing interval contains a shock at $t=0$ and it collapses at that shock at $t=\tau$. \textit{Panel (B)} The interval has no shock at $t=0$ and collapses, at $t=\tau$, at a shock which either appears within it (upper row) or merges with one of its ends (lower row) at time $t=\tstar$. In all figures, $t$, $\tstar$ and $\tau$ have been non-dimensionalized with $\TL$.}}
\end{figure*}

We define $\tfm(\ell)$ as the time taken for an interval, with initial length $\ell$, to collapse to a point at a shock, with $t=0$ the time at which we seed the flow with tracers. We consider many such intervals at $t=0$ and address the following questions: 
\begin{itemize}
\item \textit{What is the PDF of $\tfm(\ell)$?}
\item \textit{How does this PDF depend on $\ell$?} 
\end{itemize}
We start by investigating the moments of this PDF.

\subsection{Dynamic Scaling: Theoretical Considerations }\label{subsec:tcol_dynscal_Theory}
\begin{figure*}[t]
    \centering
    \includegraphics[scale=0.255]{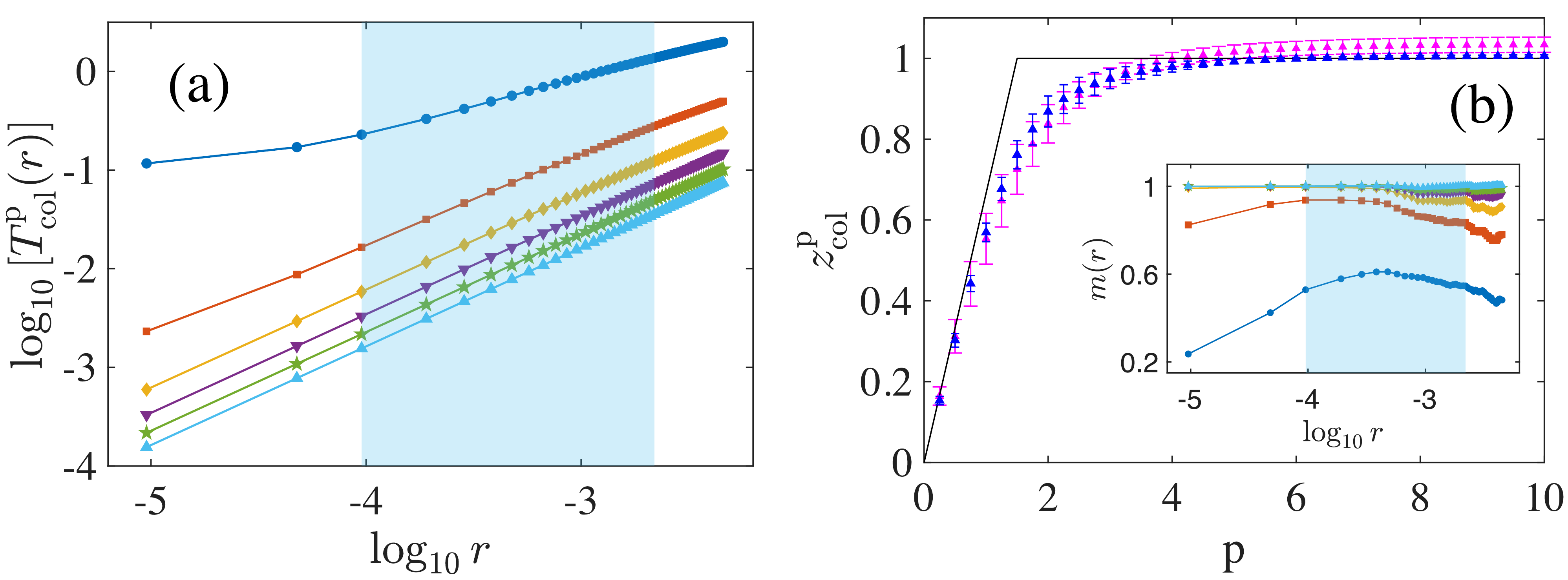}
    \caption{\small (a) Log-log plots of $\Tpfm(r)$ versus $r$ for $p=1$ (dark blue circles), $p=2$ (red squares), $p=3$ (yellow diamonds), $p=4$ (violet inverted triangles), $p=5$ (green pentagrams), and $p=6$ (light blue triangles) from run R2; we extract the exponents $\zpfm$ by calculating the local slopes of these graphs across the shaded region. (b) The exponent $\zpfm$ versus $p$ from runs R1 (pink triangles) and R2 (blue triangles). The solid line shows our theoretical prediction, Eq. \eqref{eq:zpfm2}. \textit{Inset:} The local slope $m(r)$ as a function of $r$, the color and symbol coding being the same as that in (a). The means and standard deviations of $m(r)$, calculated over the shaded region, yield the exponents $\zpfm$ and their error bars, respectively.}
    \label{fig:tfmfig}
\end{figure*}
We extract a hierarchy of time scales, $\Tpfm$, by normalizing 
order-$p$ moments of the PDF of $\tfm$ by the the large-eddy turnover time, $\TL=\Li/\urms$, with $\urms$ the root-mean-square velocity, as follows:
\begin{equation}
    \Tpfm(r)\equiv\frac{1}{\TL^{p-1}}\bra{\tfm(r,x)^p}_x \, ;
\label{eq:Tfm}
\end{equation}
$\bra{\cdot}_x$ denotes an average over different spatial locations $x$ and $r\equiv\ell/L$ is the non-dimensionalized interval length. The velocity difference, $\delta u(r)$, across inertial range separations $\eta/L\ll r\ll 1$ in a turbulent flow scales as $\delta u(r)\sim v_0r^h$, where $v_0 \equiv u_0/\urms$, the large-scale fluctuating velocity field is $u_0$, and $h$ is the scaling exponent of the velocity fluctuations. Theoretical considerations suggest~\cite{BurgMitra} that the NESS of Eqs.~(\ref{eq:Burgers}) and (\ref{eq:noise}) exhibits the following bifractal scaling, for equal-time velocity structure functions:
\begin{subequations}
\begin{align}
    \delta u(r,x) &\equiv u(x+r) - u(x) \, ; \\
    S_{\rm p}(r) &\equiv \bra{\left| \delta u(r,x)\right|^p}_x \sim r^{\zeta_{p}} \, ; \\
    \zeta_{p} &= \begin{cases}
    p/3 &\text{for}\quad p \leq 3 \, ; \\
    1 &\text{for}\quad p > 3 \, .
    \end{cases} 
\end{align}
\label{eq:eqstrfn}
\end{subequations}
This bifractal scaling is obtained as follows: For an interval of length $r$, (a) $\delta u(r,x) \sim v_0r^h$, if the interval does not contain any shock; (b) $\delta u(r,x) \sim v_0$, independent of
$r$, if it does contain a shock. The value of the exponent $h$ depends on $\beta$; for $\beta=1$, $h=1/3$. In the limit $r\ll1$, an interval of length $r$ can have at most one shock with a probability $p_r \propto r $ \footnote{This holds unless the shocks are distributed on a fractal set, which is not the case here.}. If we substitute these expressions for $\delta u(r,x)$ in $S_{\rm p}(r)$, we obtain the result for $\zeta_{p}$ in Eq.~(\ref{eq:eqstrfn}). The DNS results of Ref.~\cite{BurgMitra} yield multiscaling; but it has been suggested in this study that such multiscaling might be an artifact that should be replaced by the bifractal scaling  [Eq.~(\ref{eq:eqstrfn})] in the limit of infinite spatial resolution.

We generalize these theoretical arguments to obtain exact expressions for the dynamic scaling exponents of the NESS of Eqs.~(\ref{eq:Burgers}) and (\ref{eq:noise}). At any instant of time $t$, consider two Lagrangian particles at the two ends, $x$ and $x+y$,  of an interval of length $y$. Then 
\begin{equation}
    \dot{y}(x) = u(x+y) - u(x) \equiv \delta u(y,x)\/,
    \label{eq:drdt}
\end{equation}
where $\dot{y}=dy/dt$. The time of collapse of an interval $\tau(r,x)\equiv\tfm(r,x)/T_L$ is obtained by integrating Eq.\eq{eq:drdt}:
\begin{equation}
    \tau(r,x) \sim \int_r^0\frac{dy}{\delta u(y,x)}\, .
    \label{eq:tcolscale}
\end{equation}
In the small-$r$ limit, we have the two  
cases \textit{A} and \textit{B}, which we describe below and
illustrate in Fig. \ref{fig:tfmillus}.\\

\textit{\textbf{(A)}} There is a shock within the collapsing interval of size $r$ at $t=0$ (see panel \textit{(A)} in Fig. \ref{fig:tfmillus}). In this case $\delta u(y,x) \sim v_0$ is independent of $y$, for all $y\in[0,r]$. Hence, from Eq. \eq{eq:tcolscale} we obtain 
\begin{equation} 
\tau(r,x) \sim \frac{r}{v_0} \, .
\label{eq:tauA}
\end{equation}
The probability that an interval of size $r$ contains a shock is proportional to $r$ (this probability is $ar$, with $a$ a constant that does not affect the power laws that we obtain below), so
\begin{equation}
    \TpfmA(r) \sim r \left(\frac{r}{v_0}\right)^p \sim r^{p+1}\, ; \quad \text{case A} \, .
\label{eq:T_A}
\end{equation}\\

\textit{\textbf{(B)}} There is no shock in the interval at $t=0$ (see panel \textit{(B)} in Fig. \ref{fig:tfmillus}). As we have noted below Eq.\eq{eq:eqstrfn}, $\delta u(r,x) \sim r^h$ for this case. However, either a shock forms within the collapsing interval (first row in panel \textit{(B)} of Fig. \ref{fig:tfmillus}) or one of the ends of the interval gets trapped at a shock (second row in panel \textit{(B)} of Fig.~\ref{fig:tfmillus}), at some time $\tstar$. The probability $p_{s1}$ of finding an interval of initial size $r$, which collapses as in the first row of panel B in Fig.~\ref{fig:tfmillus}, is $p_{s1}\equiv br$, where $b$ is a constant that does not affect our results for exponents. Similarly, the probability of finding an interval, which collapses as in the second row of panel B in Fig.~\ref{fig:tfmillus}, is $p_{s2} \equiv 1-(a+b)r$. We must now distinguish between the following two sub-cases.

\textit{(B1)} $\tstar$ is very close to $\tau(r,x)$, i.e.,  $\tstar\sim\tau(r,x)$. Furthermore, $\delta u(y,x) \sim y^h$, for all $y\in[\eta,r]$, where $\eta\to 0$ in the inviscid limit; hence, by using Eq. \eq{eq:tcolscale}, we obtain
\begin{equation}
\tau(r,x) \sim \frac{1}{v_0}r^{(1-h)}\, .
\label{eq:tauB1}
\end{equation}
Hence, the contribution from this case
to $\Tpfm(r)$ is
\begin{multline}
    \TpfmBI(r) \sim w_1\tau^p p_{s1} + w_2\tau^p p_{s2}\\
    \sim w_2 r^{p(1-h)} + \mathcal{O}(r^{p(1-h)+1}) \,; \quad \text{case B1} \,.
\label{eq:T_B1}
\end{multline}
$w_1$ and $w_2$ are the weights associated with the respective types of interval collapse (see above). The first term on the right-hand side is the dominant one in the small-$r$ limit.

\textit{(B2)} $\tstar$ is significantly smaller than $\tau$, i.e., $\tstar<\tau$, so case \textit{(A)} follows after the time $\tstar$, i.e., 
\begin{equation}
    \tau(r,x) \sim \tstar + \frac{r_1}{v_0} \, ,
    \label{eq:tau_B2}
\end{equation}
where $r_1$ is the size of the interval when the shock forms. 
Then, by integrating Eq. \eq{eq:tcolscale} from $t=0$ to $\tstar$,
during which time $\delta u(y,x) \sim y^h$, we obtain
\begin{equation}
    v_0\tstar \sim \frac{1}{1-h}\left[r_1^{1-h} - r^{1-h} \right] \,.
    \label{eq:tstarB2}
\end{equation}
By solving for $r_1$ from Eq. \eq{eq:tstarB2}, substituting it in Eq. \eq{eq:tau_B2}, expanding in powers of $r$, and, in the small-$r$ limit, neglecting terms of order $r^h$
compared to unity, we obtain 
\begin{equation}
\tau(r,x) \sim \tstar + \frac{r}{v_0} \,.
\label{eq:tauB2}
\end{equation}
Hence,
\begin{equation}
    \TpfmBII(r) \sim w_3r + w_4r^p \,; \quad \text{case B2}.
\label{eq:T_B2}
\end{equation}
Here, we have incorporated the contributions from the two modes of interval collapse (cf. Eq. \eqref{eq:T_B1}), expanded the $p$-th power of $\tau$ from Eq. \eq{eq:tauB2}, and kept only the leading-order term in the small-$r$ limit; $w_3$ and $w_4$ are the weights of the respective leading-order contributions.\\ 

We combine the contributions from cases $(A)$, $(B1)$, and $(B2)$ [Eqs. \eq{eq:T_A}, \eq{eq:T_B1}, and \eq{eq:T_B2}] to obtain
\begin{equation}
     \Tpfm(r) \sim W_1 r^{p+1} + W_2 r^{p(1-h)} + W_3 r + W_4 r^p\, ,
\label{eq:Tall}
\end{equation}
where $W_1$, $W_2$, $W_3$ and $W_4$, the respective weights of these contributions, do not affect the values of the dynamic-scaling exponent, because, for any positive $p$, the second or last terms are the dominant ones ($r\ll 1$). Consequently, the dynamic-scaling 
exponents, defined by
\begin{equation}
    \Tpfm(r)\sim r^{\zpfm} \,,
    \label{eq:dynansatz}
\end{equation}
are
\begin{equation}
    \zpfm = \begin{cases}
    \frac{2p}{3} &\text{for}\quad p \leq \frac{3}{2} \, ,\\
    1 &\text{for}\quad p > \frac{3}{2} \, ,
    \end{cases}
    \label{eq:zpfm2}
\end{equation}
where we have set $h=1/3$. Thus, the bifractal scaling of equal-time velocity structure functions [Eq. \eq{eq:eqstrfn}], in the $d=1$ model of Burgers tuurbulence that we consider, is accompanied by \textit{bifractal dynamic scaling} that is embodied in Eq. \eq{eq:zpfm2}.

\subsection{Dynamic Scaling: Direct Numerical Simulations}\label{subsec:tcol_dynscal_DNS}
We now use data from our DNSs to calculate $\Tpfm(r)$; and we present log-log plots of it versus $r$, for $1 \leq p \leq 6$, in Fig. \ref{fig:tfmfig}(a). For all these values of $p$, $\Tpfm(r)$ shows power-law scaling over almost a decade and a half in $r$. From these power-law regions [in the blue-shaded rectangle in Fig. \ref{fig:tfmfig}(a)], we extract the interval-collapse exponents $\zpfm$ via a local-slope analysis of the curves. In Fig. \ref{fig:tfmfig}(b), we plot $\zpfm$ as a function of $p$, with pink and blue triangles for $\zpfm$ from runs R1 and R2, respectively. In the inset of Fig. \ref{fig:tfmfig}(b), we plot the local slopes, $m(r)$, whose means and standard deviations across the blue-shaded portion of the inset yield $\zpfm$ and its error bars, respectively. The black lines indicate the bifractal predictions of Eq. \eqref{eq:zpfm2}. Note that the values of $\zpfm$ from runs R1 and R2 lie within error bars of each other; moreover, the exponents from the higher-resolution run R2 lie slightly closer to the bifractal-scaling prediction than those from run R1. The deviation from this prediction is most pronounced near the transition value, $p=3/2$.

Our numerical results for $\zpfm$ [Fig. \ref{fig:tfmfig}(b)] indicate \textit{dynamic multiscaling}, much as the numerical results in Ref.~\cite{BurgMitra} suggest multiscaling of the equal-time velocity structure function exponents $\zeta_{p}$. However, as suggested in Ref.~\cite{BurgMitra}, it behooves us to ask whether this is \textit{bona fide} multiscaling or a numerical artifact. We remark that the small deviation of our DNS results for $\zpfm$ from the bifractal-scaling predictions [Eq. \eqref{eq:zpfm2}] are comparable to what has been observed~\cite{BurgMitra} for the equal-time velocity-structure-function exponents $\zeta_{p}$. It is possible that this small deviation could decrease as we increase the resolution of our DNS. However, given the tiny difference between the results for runs R1 and R2, we might well have to go to DNS resolutions that are computationally prohibitive before our results for $\zpfm$ approach the bifractal result of Eq. \eqref{eq:zpfm2}. Note that our DNS run R2 is, to date, the highest-resolution DNS of Eqs.  
\eqref{eq:Burgers} and \eqref{eq:noise}.

We have also checked whether the interval-collapse-time exponents $\zpfm$ depend on using different representative initial conditions, from the NESS of Eqs. \eqref{eq:Burgers} and \eqref{eq:noise}, into which we  introduce $N_p$ equi-spaced tracers to start the runs with particles. We find that, given a large-enough spatial resolution, this dependence is small, and it lies within the errors bars that we have given for $\zpfm$. We show this explicitly in Fig.~\ref{fig:manyfm} in Appendix~\ref{app:B} with four representative DNSs with $N=2^{16}$ collocation points.

\subsection{Probability Distribution Functions (PDFs) of interval-collapse times}\label{subsec:tcol_prob}
\begin{figure*}[t]
    \centering
    \includegraphics[scale=0.258]{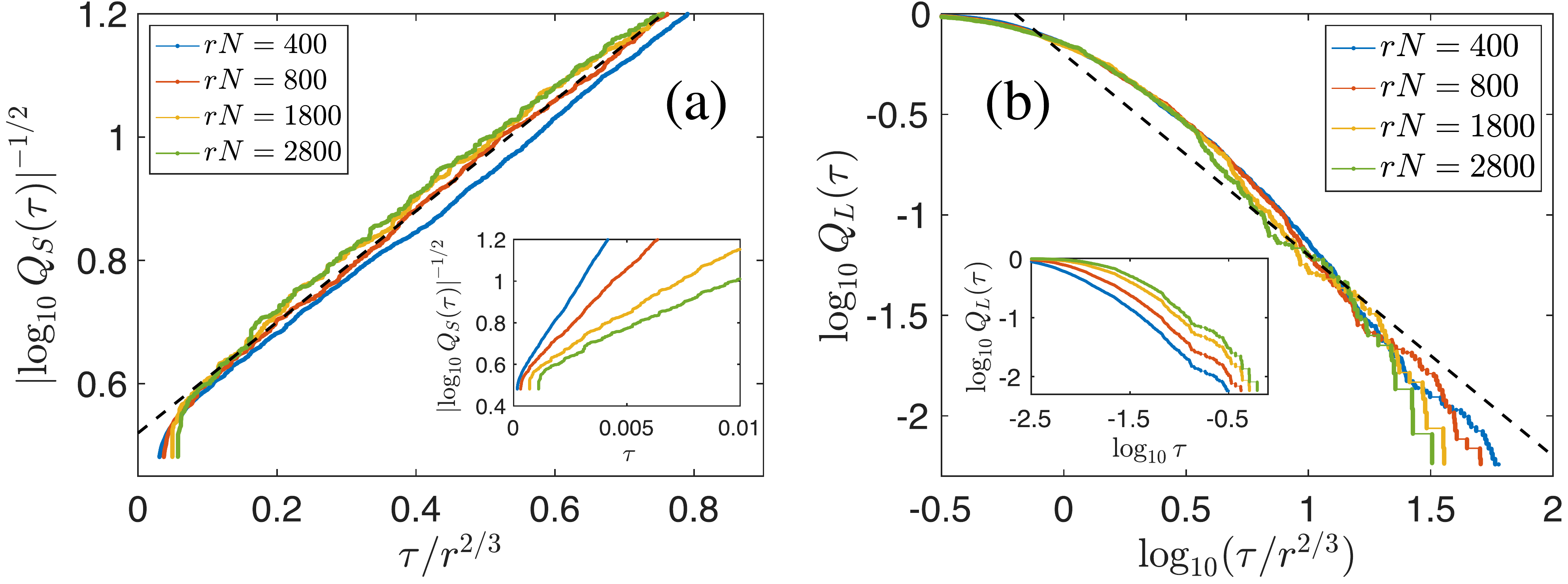}
    \caption{\small Plots for run R2: (a) $\left|\log_{10}Q_S(\tau)\right|^{-1/2}$ versus $\tau/r^{2/3}$ for different values of $rN$; the black dashed straight line represents Eq. \eqref{eq:QSscale}. \textit{Inset:}  $\left|\log_{10}Q_S(\tau)\right|^{-1/2}$ versus $\tau$ for different values of $rN$. (b) Log-log plots of the c-CPDFs, $Q_L$, versus $\tau/r^{2/3}$ for different values of $rN$; the black dashed line has a slope of $-1$. The c-CPDFs collapse, reasonably well,
    onto one curve when we scale $\tau$ by $r^{2/3}$; this curve exhibits a power-law decay with an exponent that deviates slightly from the prediction of $-1$ [see  Eq. \eqref{eq:QLscale}]. \textit{Inset:} Log-log plots of $Q_L$ versus $\tau$ for different values of $rN$.}
    \label{fig:cpdfs}
\end{figure*}
We turn now to the PDFs, $\Phi(\tfm)$, of $\tfm$ for different values of $r$; for notational convenience we suppress the argument $r$ of $\tau \equiv \tfm/T_L$ in this subsection.

\subsubsection{PDFs: Theoretical Considerations}\label{subsec:tcol_prob_Theory}
We use the relation
\begin{equation}
    \Phi(\tau)=\mathcal{P}(v_0)\left|\frac{dv_0}{d\tau}\right| \,,
    \label{eq:PDFformula}
\end{equation}
where $\mathcal{P}(v_0)$ is the PDF of $v_0$; for the Burgers equation with an external force that is limited to small $k$, this PDF is known to be a Gaussian~\cite{pdfU}. In our DNSs of Eqs.  
\eqref{eq:Burgers} and \eqref{eq:noise}, we find this PDF to be a Gaussian too (see Appendix \ref{app:C}), i.e.,
\begin{equation}
    \mathcal{P}(v_0)\sim e^{-v_0^2/2} \,.
    \label{eq:v0PDF}
\end{equation}
We now examine the forms of $\Phi(\tau)$ for the three cases \textit{(A)}, \textit{(B1)}, and \textit{(B2)} that we have considered in Sec. \ref{subsec:tcol_dynscal_Theory}.\\

\textit{\textbf{(A)}} There is a shock in the interval of size $r$ at $t=0$. The probability of finding such an interval is $p_r \propto r$. From Eq. \eqref{eq:tauA}, $v_0\sim r/\tau$. Hence, for a given value of $r$, and in
the small-$r$ limit,
\begin{equation}
    \Phi_A(\tau) = \mathcal{P}(v_0)\left|\frac{dv_0}{d\tau}\right|p_r \sim \frac{r^2}{\tau^2}\exp\left(-\frac{r^2}{2\tau^2}\right) \,.
    \label{eq:PDFcaseA}
\end{equation}\\

\textit{\textbf{(B)}} There are no shocks within the interval at $t=0$, as in Sec.~\ref{subsec:tcol_dynscal_Theory}. The interval can collapse in one of the two ways illustrated in panel \textit{(B)} of Fig.~\ref{fig:tfmillus} and can occur with probability $p_{s1}\equiv br$ or $p_{s2}\equiv 1-(a+b)r$, depending on its mode of collapse (see above). We consider the following sub-cases:

\textit{(B1)} $\tstar\sim\tau$ and $v_0\sim r^{1-h}/\tau$, so, to leading order in $r$, in the limit $r\ll1$,
\begin{equation}
    \Phi_{B1}(\tau) \sim \frac{r^{1-h}}{\tau^2}\exp\left(-\frac{r^{2(1-h)}}{2\tau^2}\right) \,.
    \label{eq:PDFcaseB1}
\end{equation}

\textit{(B2)} Here, $\tstar<\tau$ and $v_0\sim r/(\tau-\tstar)$, so, to leading order in $r$, for $r\ll1$,
\begin{multline}
    \Phi_{B2}(\tau)\\
    \sim \sum_{\tstar}\left[\frac{r}{(\tau-\tstar)^2}\exp\left\{-\frac{r^2}{2(\tau-\tstar)^2}\right\}\right]\Theta(\tau-\tstar)w(\tstar)\,,
    \label{eq:PDFcaseB2}
\end{multline}
where the summation is over all possible values of $\tstar$, with weights $w(\tstar)$, and $\Theta$, the Heaviside step function, ensures that, for a given value of $\tstar$, the contributions for all $\tau\leq \tstar$ is zero.\\

By combining Eqs. (\ref{eq:PDFcaseA})-(\ref{eq:PDFcaseB2}) and substituting $h=1/3$, we get
\begin{multline}
    \Phi(\tau)\sim \mu_1\frac{r^{2/3}}{\tau^2}\exp\left(-\frac{r^{4/3}}{2\tau^2}\right) +\\ \mu_2\frac{r^2}{\tau^2}\exp\left(-\frac{r^2}{2\tau^2}\right) + \mu_3 \Phi_{B2}(\tau) \, ,
    \label{eq:PDFtcol}
\end{multline}
where $\mu_1$, $\mu_2$, and $\mu_3$ are the weights of three contributions to $\Phi(\tau)$. We can recover the values $\zpfm$, given in Eq. (\ref{eq:zpfm2}), by calculating the moments of $\Phi(\tau)$ [Appendix \ref{app:D}], and thus check our derivation of $\Phi(\tau)$. In summary, the contributions to the order-$p$ moments of $\Phi(\tau)$,  from the first two terms on the right-hand-side (RHS) of Eq. \eqref{eq:PDFtcol}, scale as $r^{2p/3}$ and $r^{p+1}$, respectively. The leading-order contribution from the last term, $\Phi_{B2}$, scales linearly with $r$ and this comes from the binomial expansion of the term $\tau^p$ after making the substitution, $s=r/(\tau-\tstar)$.
Note that Eq. \eqref{eq:PDFtcol} explicitly demonstrates that there is no unique dynamic scaling exponent; the arguments of the two exponentials suggest dynamic exponents of $2/3$ and $1$, respectively; whereas, the power-law factors multiplying the exponentials suggest dynamic exponents of $1/3$, $1$ and $1/2$, respectively. Furthermore, this PDF is not  a Gaussian.

\subsubsection{PDFs: Direct Numerical Simulations (DNSs)}\label{subsec:tcol_prob_DNS}
We now compare the analytical form of $\Phi(\tau)$  [Eq. \eqref{eq:PDFtcol}] with the results that we obtain from our DNSs for run R2. An accurate determination of the tails of $\Phi(\tau)$ is difficult because of binning errors and insufficient sampling of rare events in these tails. Hence, we use the rank-order method~\cite{BurgMitra}, which circumvents binning errors, to calculate the small-$\tau$ and large-$\tau$ limits of the cumulative PDF (CPDF), $Q_S(\tau)$, and the complementary CPDF (c-CPDF), $Q_L(\tau)$, of $\tau$, respectively, which are defined as follows:
\begin{equation}
    Q_S(\tau)=\int_{0}^{\tau} \Phi(\tau')d\tau' \, ; \quad Q_L(\tau)=\int_{\tau}^{\infty} \Phi(\tau')d\tau'\, .
    \label{eq:Qdef}
\end{equation}
In the $\tau \to 0$ limit, the contribution from $\Phi_{B2}(\tau)$ is negligible, so, for fixed $r \ll 1$, $Q_S(\tau)$ scales as
\begin{equation}
    Q_S(\tau)\sim {\rm erfc}\left(\frac{r^{2/3}}{\tau}\right)
    \sim \frac{\tau}{r^{2/3}}\exp\left(-\frac{r^{4/3}}{\tau^2}\right)\, ,
    \label{eq:QS}
\end{equation}
where ${\rm erfc}(x)$ is the complementary error function. In the second part of Eq.~(\ref{eq:QS}), we use the leading-order asymptotic form of ${\rm erfc}(x)$ in the limit of $x\to\infty$. Furthermore, $\frac{r^{4/3}}{\tau^2}\gg\left|\log\left(\frac{\tau}{r^{2/3}}\right)\right|$, so, in the limit of $\tau\to 0$, we get
\begin{equation}
    \Big|\log Q_S(\tau)\Big|\sim \left(\frac{\tau}{r^{2/3}}\right)^{-2} \,.
    \label{eq:QSscale}
\end{equation}

In Fig. \ref{fig:cpdfs}(a), we plot $\left|\log_{10}Q_S\right|^{-1/2}$ versus $\tau/r^{2/3}$ for different values of $rN$, with $N$ the number of collocation points. In the inset, we plot $\left|\log_{10}Q_S\right|^{-1/2}$ versus $\tau$ for different values of $rN$. The plots in Fig. \ref{fig:cpdfs}(a) collapse, fairly well, onto the dashed straight line
[Eq. \eqref{eq:QSscale}], thereby providing numerical support for our $\tau \to 0$ result for $Q_S(\tau)$.

In the large-$\tau$ limit, we expand the RHS of \eqref{eq:PDFtcol} in a Taylor series, retain only the leading-order contribution for $r\ll 1$, and arrive at the following form for the c-CPDF:
\begin{equation}
    Q_L(\tau) \sim \left(\frac{\tau}{r^{2/3}}\right)^{-1} \,.
    \label{eq:QLscale}
\end{equation}

In Fig. \ref{fig:cpdfs}(b), we present log-log plots of $Q_L(\tau)$ versus $\tau/r^{2/3}$ for different values of $rN$; the inset shows log-log plots of $Q_L(\tau)$ versus $\tau$. We observe a reasonable collapse of the c-CPDFS onto a curve, which is close to, but steeper than, the straight dashed line that represents Eq. \eqref{eq:QLscale}. This discrepancy arises because of the higher-order terms, which we have neglected in
the Taylor expansion of the RHS of Eq. \eqref{eq:PDFtcol}, and which have the same sign as that of the dominant term, so we expect the curves of $Q_L$ to be steeper for moderately large values of $\tau/r^{2/3}$. The corrections because of these higher-order terms reduce on increasing $\tau$ or decreasing $r$. Consequently, in Fig. \ref{fig:cpdfs}(a), we observe that, for small values of $r$, the plots of $Q_L(\tau)$ (blue and red curves) tend to align with the dashed black line with slope $-1$, at large values of $\tau/r^{2/3}$; this alignment is not so good as $r$ increases.

\section{Time-dependent Structure Functions}\label{sec:QLstr}
\begin{figure*}[t]
\begin{centering}
\includegraphics[scale=0.257]{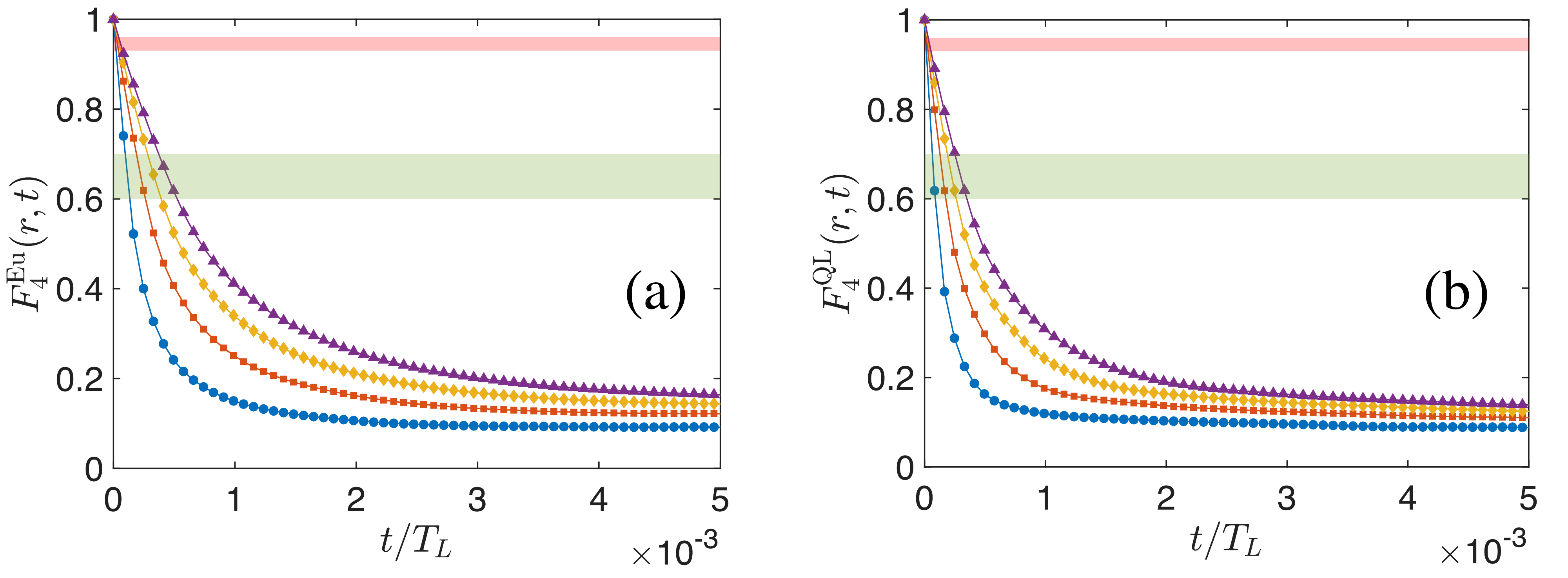}
\par\end{centering}
\caption{\label{fig:dynstrs}{\small Plots of (a) $\FEup(r,t)$ and (b) $\FQLp(r,t)$ versus $t$ for $p=4$ and $rN/L=100$ (blue circles), $200$ (red squares), $300$ (yellow diamonds) and $400$ (violet triangles). In both figures, the red and green shaded areas demarcate the regimes $0.93<\Fp(r,t)<0.96$ and $0.6<\Fp(r,t)<0.7$, respectively. Across each of these regimes, the integral-time-scale exponents remain unchanged (within numerical error bars).}}
\end{figure*}

Studies of dynamic scaling in incompressible-fluid turbulence use time-dependent structure functions~\cite{SpatStr1,IshiharaDyn,BiferaleCorr,MitraDyn,SSR_NJP,SSR_PRL,MHDShellDyn}. In some cases they employ QL structure functions to uncover dynamic multiscaling, which is masked by the sweeping effect in the Eulerian framework.
We show how to apply this QL approach to the study of  dynamic scaling in the stochastically forced $d=1$ Burgers equation [Eqs. \eqref{eq:Burgers} and \eqref{eq:noise}].

The order-$p$ \textit{time-dependent} structure function is defined as
\begin{equation}
    \Fp(r,t)\equiv\frac{1}{\Sp(r)}\left<\delta u(x,r,t_0)[\delta u(x,r,t_0+t)]^{p-1}\right>_x \,,
    \label{eq:Fp}
\end{equation}
where $\delta u(x,r,t)=\left|u(x+r,t)-u(x,t)\right|$ and $\Sp(r)$ is the order-$p$ \textit{equal-time} structure function [Eq. \eqref{eq:eqstrfn}b]. By definition, $\Fp(r,0)=1$ for all $r$. From $\Fp(r,t)$ we extract \textit{integral time scales} of orders $p$ and degree $1$, as follows:
\begin{equation}
    \Tint(r)\equiv\int_{0}^{\infty}\Fp(r,t)dt \sim r^{\chiInt}\,,
    \label{eq:intscale}
\end{equation}
where the second part defines $\chiInt$, the order-$p$ \textit{integral-time-scale exponent}, by using the \textit{dynamic scaling Ansatz}. The multifractal model of turbulence allows us to derive the following bridge relations~\cite{MitraDyn,SSR_NJP,SSR_PRL} between $\chiInt$ and the equal-time exponents $\zeta_{p}$:
\begin{equation}
    \chiInt=1+\zeta_{p-1}-\zeta_{p} \,.
    \label{eq:bridge}
\end{equation}
If we use Eq. \eqref{eq:eqstrfn}c, we obtain:
\begin{equation}
    \chiInt = \begin{cases}
    \frac{2}{3} &\text{for}\quad p < 3 \,; \\
    \frac{p-1}{3} &\text{for}\quad 3\leq p\leq 4 \,; \\
    1 &\text{for}\quad p > 4 \,.
    \end{cases}
    \label{eq:chip}
\end{equation}
\begin{figure*}[t]
    \centering
    \includegraphics[scale=0.26]{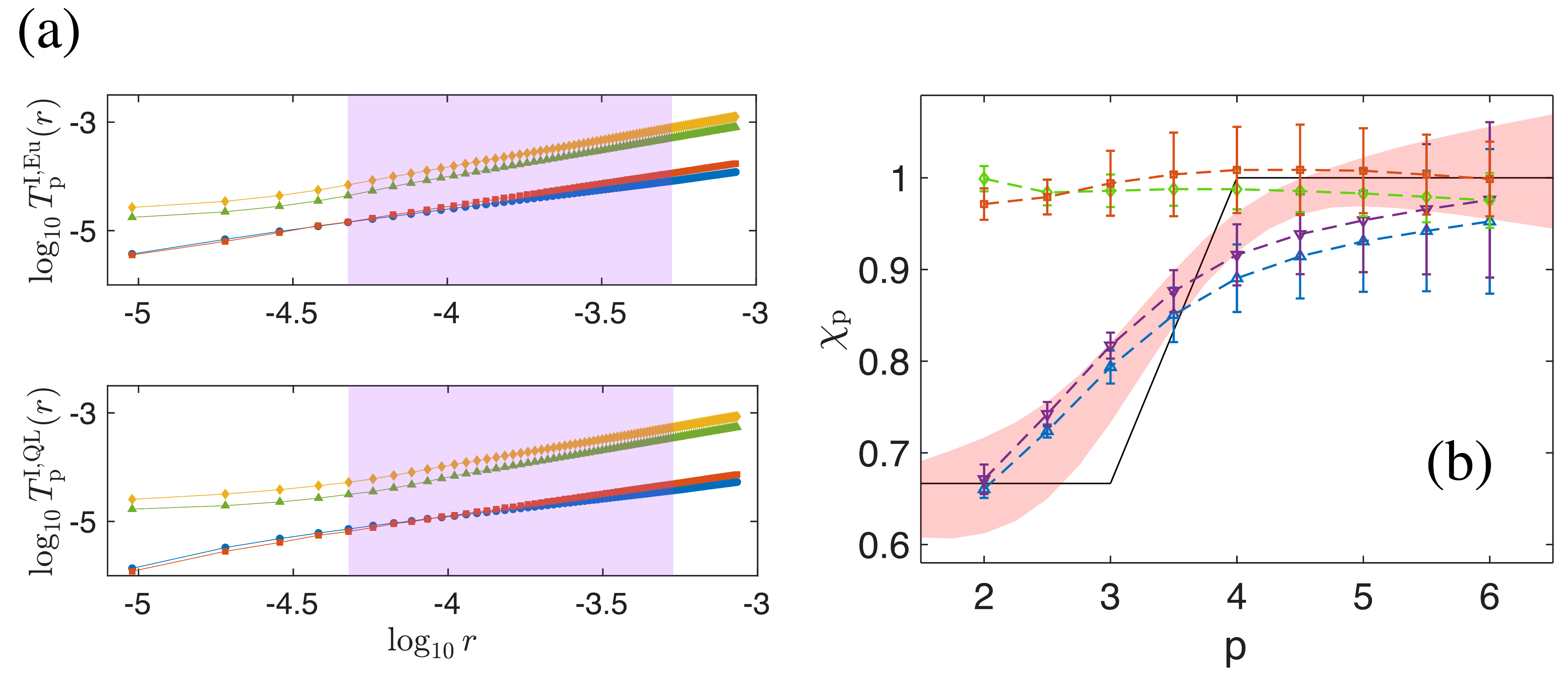}
    \caption{\small Plots for run R2: (a) Log-log plots of the integral time scales, $\TintEu(r)$ and $\TintQL(r)$, versus $r$. In both panels, the blue circles and red squares correspond to orders $p=2$ and $p=3$, respectively, with $\lambda=0.95$ (see text); points obtained with $\lambda=0.65$ are indicted by  yellow diamonds and green triangles for orders $p=2$ and $p=3$, respectively. The pink shaded area indicates the regime in which we carry out local-slope analyses. (b) Blue and violet triangles are the integral-time-scale exponents, $\chi^{I,\,QL}_p$ and $\chi^{I,\,Eu}_p$, respectively, calculated with $\lambda=0.95$; the shaded area is the QL-bridge-relation prediction obtained by substituting $\zetaQL_{\rm p}$ (see Appendix \ref{app:E}) in Eq. \eqref{eq:bridge}; the black lines are the bifractal prediction, Eq. \eqref{eq:chip}; the red squares and green diamonds represent $\chi^{I,\,QL}_p$ and $\chi^{I,\,Eu}_p$, respectively, calculated with $\lambda=0.65$. The Eulerian and QL bridge-relation predictions are equal up to three decimal places.}
    \label{fig:Qlexpfig}
\end{figure*}

We calculate $\Sp(r)$, $\Fp(r,t)$ and $\Tint(r)$ from our DNS by using both the Eulerian velocity $u(x,t)$ and the QL velocity 
\begin{equation}
    V(x,t)=u(x+R(t),t) \,,
    \label{eq:QL}
\end{equation}
where $R(t)$ is the position of a Lagrangian particle, at time $t$, which was at position $x_0$ at $t=0$ (see Appendix  \ref{app:A} for details). Henceforth, we denote the Eulerian structure functions, time scales, and scaling exponents with the superscript Eu (i.e., $\FEup$, $\SEup$, $\TintEu$, $\zetaEu_{\rm p}$, $\chiEu$) and their QL counterparts with the superscript QL ($\FQLp$, $\SQLp$, $\TintQL$, $\zetaQL_{\rm p}$, $\chiQL$). In Fig. \ref{fig:dynstrs}, we plot $\FEup$ and $\FQLp$ versus $t$, for different values of $r$, for the order $p=4$.

We extract the Eulerian integral time scales, $\TintEu(r)$, from Eq. \eqref{eq:intscale} by integrating $\FEup(r,t)$ via the trapezoidal rule. The larger the time delay $t$, the more unreliable are the numerical data for these structure functions. Therefore, we use a time $\taud$ as the upper limit of the integral, with $\taud$ the time at which $\FEup$ decreases to a certain fraction $\lambda$. We consider both the short- and late-time estimates for $\TintEu(r)$ by employing large and small values of $\lambda$, respectively.

For $\lambda=0.95$ we plot $\TintEu(r)$ for different values of $p$ in the upper panel of Fig. \ref{fig:Qlexpfig}(a); and from the local slopes of these graphs, we evaluate the dynamic exponents $\chiEu$ and display them via blue triangles in Fig. \ref{fig:Qlexpfig}(b). These exponents closely match with the multifractal-bridge-relation prediction, which we derive by substituting the Eulerian equal-time structure function exponents $\zetaEu_{p}$ (see Appendix \ref{app:E}) into Eq. \eqref{eq:bridge} and which is indicated by the shaded region in Fig. \ref{fig:Qlexpfig}(b). The width of this shaded region indicates the error bars in the equal-time exponents that are used in the bridge relation. The values of $\chiEu$ remain unchanged, within error bars, for $0.93<\lambda<0.96$.

Next we calculate the integral time scales with $\lambda=0.65$. We observe that the corresponding dynamic exponents $\chiEu$ [green diamonds in Fig. \ref{fig:Qlexpfig}(b)] lie close to, and within error bars of, unity for all $p$. These results remain unchanged for $0.6<\lambda<0.7$.

Similar Eulerian results were obtained previously by Hayot and Jayaprakash~\cite{HayotDynStr}. By using a different class of time-dependent structure functions, they defined a dynamic exponent $z$ and observed that $z$ is indeed unity at intermediate time scales, whereas, at short times, it is not unity. They suggested that this was a manifestation of the Taylor hypothesis or the sweeping effect: The advection of the small eddies by large ones yields a linear relation between spatial and temporal scales and thence a dynamic exponent $z=1$.

We now calculate the QL structure functions $\FQLp$,  extract $\TintQL(r)$ [see the lower panel of Fig. \ref{fig:Qlexpfig}(a)], and thence the exponents $\chiQL$, in the same way as we did above for their Eulerian counterparts. In particular, we use (a) $0.93<\lambda<0.96$ and (b) $0.6<\lambda<0.7$. In case (a), $\chiQL$ [violet inverted triangles in Fig. \ref{fig:Qlexpfig}(b)] agrees with the bridge-relation predictions of Eq. \eqref{eq:bridge}, derived from $\zetaQL_{p}$.
In case (b), we find that $\chiQL$ lies within error bars of unity for all $p$ [red squares in Fig. \ref{fig:Qlexpfig}(b)], like the corresponding Eulerian exponents.

Thus, QL structure functions, which eliminate sweeping effects when we study dynamic multiscaling in incompressible turbulence~\cite{SpatStr1,IshiharaDyn,BiferaleCorr,MitraDyn,SSR_NJP,SSR_PRL,MHDShellDyn},  do not remove sweeping effects in the $d=1$ Burgers turbulence that we study. The qualitative reason for this is that tracers get trapped in shocks and, thereafter, they move with the shock. We conjecture that this inability of QL structure functions to remove sweeping effects holds in all models of compressible turbulence, once shocks are formed. 

\section{Conclusion and Discussions}\label{sec:Discussions}

We have carried out a detailed study of dynamic multiscaling in the NESS of the stochastically forced $d=1$ Burgers equation [Eqs. \eqref{eq:Burgers} and \eqref{eq:noise}]. In particular, we have shown that there is not one but an infinite number of time-scales, associated with a single length scale, so this NESS displays dynamic multiscaling. We have proposed a theoretical framework to calculate these exponents. As mentioned earlier, to the best of our knowledge, \textit{this is the first time that such dynamic multiscaling exponents have been obtained analytically for turbulence in any nonlinear hydrodynamical PDE.} We have validated our theoretical findings with the results of our high-resolution direct numerical simulations. We have also quantified the behaviors of the the tails of the CPDFs of $\tfm$ for different $\ell$. Furthermore, we have explicitly shown that the QL methods~\cite{SpatStr1,BiferaleCorr,MitraDyn,SSR_NJP,SSR_PRL,MHDShellDyn}, which have been used to study such multiscaling of turbulence in incompressible flows are not suitable for Burgers turbulence because it is compressible.

Our DNS results for $\zpfm$ deviate slightly from the bifractal prediction [Eq. \eqref{eq:zpfm2}], which seems to indicate dynamic multiscaling, much like what is observed in Ref.~\cite{BurgMitra} for the equal-time exponents $\zeta_{p}$. As suggested in Ref.~\cite{BurgMitra}, this deviation might possibly be a numerical artifact, rather than \textit{bona fide} dynamic multiscaling, and could well decrease on increasing the resolution of our DNS by a few orders of magnitude.

It should be possible to use the multifractal model of turbulence  to derive a bridge relation that links the collapse-time exponents $\zpfm$ to the equal-time structure function exponents $\zeta_{p}$. The multifractal model has been used, earlier, to derive the integral-time bridge relation Eq. \eqref{eq:bridge}, from Eq. \eqref{eq:intscale} in incompressible fluid turbulence~\cite{SpatStr1,IshiharaDyn,BiferaleCorr,MitraDyn,SSR_NJP,SSR_PRL,MHDShellDyn}. We outline a similar derivation for $\zpfm$ in our $d=1$ Burgers turbulence model in Appendix \ref{app:F}, where we also discuss its limitations.

To generalize our study to the $d-$dimensional case, we consider the stochastically forced Burgers equation
\begin{equation}
    \partial_t{\bf u}+({\bf u}\cdot\nabla){\bf u} = \nu\nabla^2{\bf u}+{\bf f}({\bf x},\,t) \,,
    \label{eq:ddimBurg}
\end{equation}
with the zero-mean, white-in-time stochastic force ${\bf f}({\bf x},\,t)$. The covariance of its Fourier components satisfies
\begin{equation}
    \langle \hat{f_i}(\mathbf{k},\,t)\hat{f_j}(\mathbf{k}',\,t')\rangle\sim k^{-\beta}\delta(\mathbf{k}+\mathbf{k}')\delta(t-t')\delta_{ij} \,.
    \label{eq:ddimforc}
\end{equation}
The subscripts $i$ and $j$ denote the Cartesian components of the force. This force must be chosen to be curl free to ensure that the flow is not vortical. The choice $\beta = d$ yields a K41-type inertial-range energy spectrum at the level of a one-loop RG. In the turbulent NESS for this model [Eqs. \eqref{eq:ddimBurg} and \eqref{eq:ddimforc}] shocks are extended structures (lines if $d=2$ and surfaces if $d=3$). For introductions to such shocks, we refer the reader to Refs.~\cite{Bec2007BurgersT,Perestroika1}. The natural generalization of the interval-collapse times, which we have used above for $d=1$ Burgers turbulence, are the collapse times of Lagrangian $d-$simplices, i.e., triangles and  tetrahedra in $d=2$ and $d=3$, respectively, with tracers at their vertices. We have carried out preliminary DNSs  to study dynamic multiscaling in the NESS of the $d=2$ stochastically forced Burgers equation. We find that we must distinguish between the collapse of a triangle to one with vanishing area, in one of the following ways: \textit{(i)} all three vertices collapse to a point; \textit{(ii}) any two vertices come together; or \textit{(iii)} all three vertices become colinear, with none of them overlapping. We have verified that the collapse times for cases \textit{(i)}, \textit{(ii)}, and \textit{(iii)} capture the dynamic multiscaling of the NESS  of Eqs. \eqref{eq:ddimBurg} and \eqref{eq:ddimforc}, but with dynamic exponents that depend on which of the cases [\textit{(i)}-\textit{(iii)}] we consider. The collapse of Lagrangian tetrahedra in $d=3$  requires even more cases than that for triangles in $d=2$. The elucidation of the dependence of these exponents on initial conditions in this NESS is a challenge because it requires high-resolution DNSs (for $d=1$ Burgers turbulence see Fig.~\ref{fig:manyfm} in Appendix~\ref{app:B}). The details of our studies of dynamic multiscaling in the NESSs of Eqs. \eqref{eq:ddimBurg} and \eqref{eq:ddimforc}, for $d=2$ and $d=3$, will be given elsewhere.

Our study has obvious implications  for the elucidations of dynamic scaling in compressible turbulence (CT) with shocks. We will explore this in future work. We note that the statistics of velocity fluctuations in the stochastically  forced $d=1$ Burgers equation that we consider is similar to that in the fully detailed $d=1$ model of CT with unit Mach number, investigated in Ref.~\cite{1Dcomp}. Of course, the study of CT with shocks is extremely relevant in the engineering of combustion systems and supersonic jets and in astrophysical studies of, e.g., the solar wind, interstellar gases, and molecular clouds~\cite{Compress1,Compress2,Compress3,Compress4,Compress5,Compress6,CompressLag1,CompressLag2,CompressPl1,CompressPl2,CompressAstro1,CompressAstro2,CompressAstro3}. The application of our interval-collapse times, and their generalizations to $d >1$ hold promise for studies of the spatiotemporal statistics of the compressible turbulent flows mentioned above and those that are obtained in DNSs of compressible hydrodynamical equations. It would be interesting to explore whether times such as $\tfm$,
and their generalization to heavy inertial particles, could be related to characteristic times in processes that require the agglomeration of masses or chemical species advected by highly compressible turbulent flows 
in astrophysical turbulence (e.g., dust in molecular clouds) and in turbulent combustion~\cite{kruger2017effect,kruger2017correlation}. In fact, in Refs.~\cite{kruger2017effect,kruger2017correlation}, the authors have looked at the possible effects of clustering in weakly compressible flows. Similar studies for clustering because of shocks will be very interesting.

It would also be fascinating to explore, in detail, the relation of our work with statistical-mechanical studies, carried out in the context of the KPZ equation, which investigate the behavior of passive random walkers following the wrinkles of a rough surface whose fluctuations are determined by this equation~\cite{PRW1,PRW2,PRW3,PRW4}. 

\par\bigskip\noindent
\textit{Acknowledgments\,:} SD thanks the Prime Minister's Research Fellowship (PMRF) for support; RP thanks SERB (India), the National Supercomputing Mission (India), and SERC (IISc) for support and computational resources. DM acknowledges the support of the Swedish Research
Council Grant No. 638-2013-9243 and 2016-05225.

\bibliographystyle{apsrev4-1}
\bibliography{references.bib}

\begin{thebibliography}{72}%
\makeatletter
\providecommand \@ifxundefined [1]{%
 \@ifx{#1\undefined}
}%
\providecommand \@ifnum [1]{%
 \ifnum #1\expandafter \@firstoftwo
 \else \expandafter \@secondoftwo
 \fi
}%
\providecommand \@ifx [1]{%
 \ifx #1\expandafter \@firstoftwo
 \else \expandafter \@secondoftwo
 \fi
}%
\providecommand \natexlab [1]{#1}%
\providecommand \enquote  [1]{``#1''}%
\providecommand \bibnamefont  [1]{#1}%
\providecommand \bibfnamefont [1]{#1}%
\providecommand \citenamefont [1]{#1}%
\providecommand \href@noop [0]{\@secondoftwo}%
\providecommand \href [0]{\begingroup \@sanitize@url \@href}%
\providecommand \@href[1]{\@@startlink{#1}\@@href}%
\providecommand \@@href[1]{\endgroup#1\@@endlink}%
\providecommand \@sanitize@url [0]{\catcode `\\12\catcode `\$12\catcode
  `\&12\catcode `\#12\catcode `\^12\catcode `\_12\catcode `\%12\relax}%
\providecommand \@@startlink[1]{}%
\providecommand \@@endlink[0]{}%
\providecommand \url  [0]{\begingroup\@sanitize@url \@url }%
\providecommand \@url [1]{\endgroup\@href {#1}{\urlprefix }}%
\providecommand \urlprefix  [0]{URL }%
\providecommand \Eprint [0]{\href }%
\providecommand \doibase [0]{http://dx.doi.org/}%
\providecommand \selectlanguage [0]{\@gobble}%
\providecommand \bibinfo  [0]{\@secondoftwo}%
\providecommand \bibfield  [0]{\@secondoftwo}%
\providecommand \translation [1]{[#1]}%
\providecommand \BibitemOpen [0]{}%
\providecommand \bibitemStop [0]{}%
\providecommand \bibitemNoStop [0]{.\EOS\space}%
\providecommand \EOS [0]{\spacefactor3000\relax}%
\providecommand \BibitemShut  [1]{\csname bibitem#1\endcsname}%
\let\auto@bib@innerbib\@empty
\bibitem [{\citenamefont {Frisch}(1996)}]{frisch}%
  \BibitemOpen
  \bibfield  {author} {\bibinfo {author} {\bibfnamefont {U.}~\bibnamefont
  {Frisch}},\ }\href@noop {} {\emph {\bibinfo {title} {Turbulence the Legacy of
  A.N. Kolmogorov}}}\ (\bibinfo  {publisher} {Cambridge University Press,
  Cambridge, England},\ \bibinfo {year} {1996})\BibitemShut {NoStop}%
\bibitem [{\citenamefont {Hayot}\ and\ \citenamefont
  {Jayaprakash}(1998)}]{HayotDynStr}%
  \BibitemOpen
  \bibfield  {author} {\bibinfo {author} {\bibfnamefont {F.}~\bibnamefont
  {Hayot}}\ and\ \bibinfo {author} {\bibfnamefont {C.}~\bibnamefont
  {Jayaprakash}},\ }\href {\doibase 10.1103/PhysRevE.57.R4867} {\bibfield
  {journal} {\bibinfo  {journal} {Phys. Rev. E}\ }\textbf {\bibinfo {volume}
  {57}},\ \bibinfo {pages} {R4867} (\bibinfo {year} {1998})}\BibitemShut
  {NoStop}%
\bibitem [{\citenamefont {Hayot}\ and\ \citenamefont
  {Jayaprakash}(2000)}]{HayotPhysFlu}%
  \BibitemOpen
  \bibfield  {author} {\bibinfo {author} {\bibfnamefont {F.}~\bibnamefont
  {Hayot}}\ and\ \bibinfo {author} {\bibfnamefont {C.}~\bibnamefont
  {Jayaprakash}},\ }\href {\doibase 10.1063/1.870311} {\bibfield  {journal}
  {\bibinfo  {journal} {Phys. Fluids}\ }\textbf {\bibinfo {volume} {12}},\
  \bibinfo {pages} {327} (\bibinfo {year} {2000})}\BibitemShut {NoStop}%
\bibitem [{\citenamefont {L'vov}\ \emph {et~al.}(1997)\citenamefont {L'vov},
  \citenamefont {Podivilov},\ and\ \citenamefont {Procaccia}}]{SpatStr1}%
  \BibitemOpen
  \bibfield  {author} {\bibinfo {author} {\bibfnamefont {V.~S.}\ \bibnamefont
  {L'vov}}, \bibinfo {author} {\bibfnamefont {E.}~\bibnamefont {Podivilov}}, \
  and\ \bibinfo {author} {\bibfnamefont {I.}~\bibnamefont {Procaccia}},\ }\href
  {\doibase 10.1103/PhysRevE.55.7030} {\bibfield  {journal} {\bibinfo
  {journal} {Phys. Rev. E}\ }\textbf {\bibinfo {volume} {55}},\ \bibinfo
  {pages} {7030} (\bibinfo {year} {1997})}\BibitemShut {NoStop}%
\bibitem [{\citenamefont {Mitra}\ and\ \citenamefont
  {Pandit}(2004)}]{MitraDyn}%
  \BibitemOpen
  \bibfield  {author} {\bibinfo {author} {\bibfnamefont {D.}~\bibnamefont
  {Mitra}}\ and\ \bibinfo {author} {\bibfnamefont {R.}~\bibnamefont {Pandit}},\
  }\href {\doibase 10.1103/PhysRevLett.93.024501} {\bibfield  {journal}
  {\bibinfo  {journal} {Phys. Rev. Lett.}\ }\textbf {\bibinfo {volume} {93}},\
  \bibinfo {pages} {024501} (\bibinfo {year} {2004})}\BibitemShut {NoStop}%
\bibitem [{\citenamefont {Ray}\ \emph {et~al.}(2008)\citenamefont {Ray},
  \citenamefont {Mitra},\ and\ \citenamefont {Pandit}}]{SSR_NJP}%
  \BibitemOpen
  \bibfield  {author} {\bibinfo {author} {\bibfnamefont {S.~S.}\ \bibnamefont
  {Ray}}, \bibinfo {author} {\bibfnamefont {D.}~\bibnamefont {Mitra}}, \ and\
  \bibinfo {author} {\bibfnamefont {R.}~\bibnamefont {Pandit}},\ }\href
  {\doibase 10.1088/1367-2630/10/3/033003} {\bibfield  {journal} {\bibinfo
  {journal} {New J. Phys.}\ }\textbf {\bibinfo {volume} {10}},\ \bibinfo
  {pages} {033003} (\bibinfo {year} {2008})}\BibitemShut {NoStop}%
\bibitem [{\citenamefont {Ray}\ \emph {et~al.}(2011)\citenamefont {Ray},
  \citenamefont {Mitra}, \citenamefont {Perlekar},\ and\ \citenamefont
  {Pandit}}]{SSR_PRL}%
  \BibitemOpen
  \bibfield  {author} {\bibinfo {author} {\bibfnamefont {S.~S.}\ \bibnamefont
  {Ray}}, \bibinfo {author} {\bibfnamefont {D.}~\bibnamefont {Mitra}}, \bibinfo
  {author} {\bibfnamefont {P.}~\bibnamefont {Perlekar}}, \ and\ \bibinfo
  {author} {\bibfnamefont {R.}~\bibnamefont {Pandit}},\ }\href {\doibase
  10.1103/PhysRevLett.107.184503} {\bibfield  {journal} {\bibinfo  {journal}
  {Phys. Rev. Lett.}\ }\textbf {\bibinfo {volume} {107}},\ \bibinfo {pages}
  {184503} (\bibinfo {year} {2011})}\BibitemShut {NoStop}%
\bibitem [{\citenamefont {Pandit}\ \emph {et~al.}(2008)\citenamefont {Pandit},
  \citenamefont {Ray},\ and\ \citenamefont {Mitra}}]{PanditDynRev}%
  \BibitemOpen
  \bibfield  {author} {\bibinfo {author} {\bibfnamefont {R.}~\bibnamefont
  {Pandit}}, \bibinfo {author} {\bibfnamefont {S.}~\bibnamefont {Ray}}, \ and\
  \bibinfo {author} {\bibfnamefont {D.}~\bibnamefont {Mitra}},\ }\href
  {\doibase 10.1140/epjb/e2008-00048-6} {\bibfield  {journal} {\bibinfo
  {journal} {Eur. Phys. J. B}\ }\textbf {\bibinfo {volume} {64}},\ \bibinfo
  {pages} {463} (\bibinfo {year} {2008})}\BibitemShut {NoStop}%
\bibitem [{\citenamefont {Chaikin}\ and\ \citenamefont
  {Lubensky}(1995)}]{chaikin}%
  \BibitemOpen
  \bibfield  {author} {\bibinfo {author} {\bibfnamefont {P.~M.}\ \bibnamefont
  {Chaikin}}\ and\ \bibinfo {author} {\bibfnamefont {T.~C.}\ \bibnamefont
  {Lubensky}},\ }\href {\doibase 10.1017/CBO9780511813467} {\emph {\bibinfo
  {title} {Principles of Condensed Matter Physics}}}\ (\bibinfo  {publisher}
  {Cambridge University Press},\ \bibinfo {year} {1995})\BibitemShut {NoStop}%
\bibitem [{\citenamefont {Kardar}(2007)}]{kardar2007statistical}%
  \BibitemOpen
  \bibfield  {author} {\bibinfo {author} {\bibfnamefont {M.}~\bibnamefont
  {Kardar}},\ }\href@noop {} {\emph {\bibinfo {title} {Statistical physics of
  fields}}}\ (\bibinfo  {publisher} {Cambridge University Press},\ \bibinfo
  {year} {2007})\BibitemShut {NoStop}%
\bibitem [{\citenamefont {Goldenfeld}(2018)}]{goldenfeld2018lectures}%
  \BibitemOpen
  \bibfield  {author} {\bibinfo {author} {\bibfnamefont {N.}~\bibnamefont
  {Goldenfeld}},\ }\href@noop {} {\emph {\bibinfo {title} {Lectures on phase
  transitions and the renormalization group}}}\ (\bibinfo  {publisher} {CRC
  Press},\ \bibinfo {year} {2018})\BibitemShut {NoStop}%
\bibitem [{\citenamefont {Forster}\ \emph {et~al.}(1977)\citenamefont
  {Forster}, \citenamefont {Nelson},\ and\ \citenamefont
  {Stephen}}]{forster1977large}%
  \BibitemOpen
  \bibfield  {author} {\bibinfo {author} {\bibfnamefont {D.}~\bibnamefont
  {Forster}}, \bibinfo {author} {\bibfnamefont {D.~R.}\ \bibnamefont {Nelson}},
  \ and\ \bibinfo {author} {\bibfnamefont {M.~J.}\ \bibnamefont {Stephen}},\
  }\href@noop {} {\bibfield  {journal} {\bibinfo  {journal} {Physical Review
  A}\ }\textbf {\bibinfo {volume} {16}},\ \bibinfo {pages} {732} (\bibinfo
  {year} {1977})}\BibitemShut {NoStop}%
\bibitem [{\citenamefont {DeDominicis}\ and\ \citenamefont
  {Martin}(1979)}]{dedominicis1979energy}%
  \BibitemOpen
  \bibfield  {author} {\bibinfo {author} {\bibfnamefont {C.}~\bibnamefont
  {DeDominicis}}\ and\ \bibinfo {author} {\bibfnamefont {P.}~\bibnamefont
  {Martin}},\ }\href@noop {} {\bibfield  {journal} {\bibinfo  {journal}
  {Physical Review A}\ }\textbf {\bibinfo {volume} {19}},\ \bibinfo {pages}
  {419} (\bibinfo {year} {1979})}\BibitemShut {NoStop}%
\bibitem [{\citenamefont {Pandit}\ \emph {et~al.}(2009)\citenamefont {Pandit},
  \citenamefont {Perlekar},\ and\ \citenamefont {Ray}}]{pandit2009statistical}%
  \BibitemOpen
  \bibfield  {author} {\bibinfo {author} {\bibfnamefont {R.}~\bibnamefont
  {Pandit}}, \bibinfo {author} {\bibfnamefont {P.}~\bibnamefont {Perlekar}}, \
  and\ \bibinfo {author} {\bibfnamefont {S.~S.}\ \bibnamefont {Ray}},\
  }\href@noop {} {\bibfield  {journal} {\bibinfo  {journal} {Pramana}\ }\textbf
  {\bibinfo {volume} {73}},\ \bibinfo {pages} {157} (\bibinfo {year}
  {2009})}\BibitemShut {NoStop}%
\bibitem [{\citenamefont {Parisi}\ and\ \citenamefont
  {Frisch}(1985)}]{parisi1985multifractal}%
  \BibitemOpen
  \bibfield  {author} {\bibinfo {author} {\bibfnamefont {G.}~\bibnamefont
  {Parisi}}\ and\ \bibinfo {author} {\bibfnamefont {U.}~\bibnamefont
  {Frisch}},\ }\href@noop {} {\bibfield  {journal} {\bibinfo  {journal}
  {Turbulence and predictability in geophysical fluid dynamics and climate
  dynamics, Varenna, 1983, M. Ghil, R. Benzi and G. Parisi, eds. Proceedings of
  the International School of Physic Enrico Fermi, North-Holland 1985}\ ,\
  \bibinfo {pages} {84}} (\bibinfo {year} {1985})}\BibitemShut {NoStop}%
\bibitem [{\citenamefont {Benzi}\ \emph {et~al.}(1984)\citenamefont {Benzi},
  \citenamefont {Paladin}, \citenamefont {Parisi},\ and\ \citenamefont
  {Vulpiani}}]{benzi1984multifractal}%
  \BibitemOpen
  \bibfield  {author} {\bibinfo {author} {\bibfnamefont {R.}~\bibnamefont
  {Benzi}}, \bibinfo {author} {\bibfnamefont {G.}~\bibnamefont {Paladin}},
  \bibinfo {author} {\bibfnamefont {G.}~\bibnamefont {Parisi}}, \ and\ \bibinfo
  {author} {\bibfnamefont {A.}~\bibnamefont {Vulpiani}},\ }\href@noop {}
  {\bibfield  {journal} {\bibinfo  {journal} {Journal of Physics A:
  Mathematical and General}\ }\textbf {\bibinfo {volume} {17}},\ \bibinfo
  {pages} {3521} (\bibinfo {year} {1984})}\BibitemShut {NoStop}%
\bibitem [{\citenamefont {Halperin}\ and\ \citenamefont
  {Hohenberg}(1967)}]{halperin1967generalization}%
  \BibitemOpen
  \bibfield  {author} {\bibinfo {author} {\bibfnamefont {B.~I.}\ \bibnamefont
  {Halperin}}\ and\ \bibinfo {author} {\bibfnamefont {P.}~\bibnamefont
  {Hohenberg}},\ }\href@noop {} {\bibfield  {journal} {\bibinfo  {journal}
  {Physical Review Letters}\ }\textbf {\bibinfo {volume} {19}},\ \bibinfo
  {pages} {700} (\bibinfo {year} {1967})}\BibitemShut {NoStop}%
\bibitem [{\citenamefont {Ferrell}\ \emph {et~al.}(1968)\citenamefont
  {Ferrell}, \citenamefont {Menyhard}, \citenamefont {Schmidt}, \citenamefont
  {Schwabl},\ and\ \citenamefont {Szepfalusy}}]{ferrell1968fluctuations}%
  \BibitemOpen
  \bibfield  {author} {\bibinfo {author} {\bibfnamefont {R.}~\bibnamefont
  {Ferrell}}, \bibinfo {author} {\bibfnamefont {N.}~\bibnamefont {Menyhard}},
  \bibinfo {author} {\bibfnamefont {H.}~\bibnamefont {Schmidt}}, \bibinfo
  {author} {\bibfnamefont {F.}~\bibnamefont {Schwabl}}, \ and\ \bibinfo
  {author} {\bibfnamefont {P.}~\bibnamefont {Szepfalusy}},\ }\href@noop {}
  {\bibfield  {journal} {\bibinfo  {journal} {Annals of Physics}\ }\textbf
  {\bibinfo {volume} {47}},\ \bibinfo {pages} {565} (\bibinfo {year}
  {1968})}\BibitemShut {NoStop}%
\bibitem [{\citenamefont {Hohenberg}\ and\ \citenamefont
  {Halperin}(1977)}]{hohenberg1977theory}%
  \BibitemOpen
  \bibfield  {author} {\bibinfo {author} {\bibfnamefont {P.~C.}\ \bibnamefont
  {Hohenberg}}\ and\ \bibinfo {author} {\bibfnamefont {B.~I.}\ \bibnamefont
  {Halperin}},\ }\href@noop {} {\bibfield  {journal} {\bibinfo  {journal}
  {Reviews of Modern Physics}\ }\textbf {\bibinfo {volume} {49}},\ \bibinfo
  {pages} {435} (\bibinfo {year} {1977})}\BibitemShut {NoStop}%
\bibitem [{\citenamefont {Ray}\ \emph {et~al.}(2016)\citenamefont {Ray},
  \citenamefont {Sahoo},\ and\ \citenamefont {Pandit}}]{MHDShellDyn}%
  \BibitemOpen
  \bibfield  {author} {\bibinfo {author} {\bibfnamefont {S.~S.}\ \bibnamefont
  {Ray}}, \bibinfo {author} {\bibfnamefont {G.}~\bibnamefont {Sahoo}}, \ and\
  \bibinfo {author} {\bibfnamefont {R.}~\bibnamefont {Pandit}},\ }\href
  {\doibase 10.1103/PhysRevE.94.053101} {\bibfield  {journal} {\bibinfo
  {journal} {Phys. Rev. E}\ }\textbf {\bibinfo {volume} {94}},\ \bibinfo
  {pages} {053101} (\bibinfo {year} {2016})}\BibitemShut {NoStop}%
\bibitem [{\citenamefont {Frisch}\ and\ \citenamefont
  {Bec}(2001)}]{frisch2001burgulence}%
  \BibitemOpen
  \bibfield  {author} {\bibinfo {author} {\bibfnamefont {U.}~\bibnamefont
  {Frisch}}\ and\ \bibinfo {author} {\bibfnamefont {J.}~\bibnamefont {Bec}},\
  }in\ \href@noop {} {\emph {\bibinfo {booktitle} {New trends in turbulence
  Turbulence: nouveaux aspects}}}\ (\bibinfo  {publisher} {Springer},\ \bibinfo
  {year} {2001})\ pp.\ \bibinfo {pages} {341--383}\BibitemShut {NoStop}%
\bibitem [{\citenamefont {Bec}\ and\ \citenamefont
  {Khanin}(2007)}]{Bec2007BurgersT}%
  \BibitemOpen
  \bibfield  {author} {\bibinfo {author} {\bibfnamefont {J.}~\bibnamefont
  {Bec}}\ and\ \bibinfo {author} {\bibfnamefont {K.}~\bibnamefont {Khanin}},\
  }\href {\doibase 10.1016/j.physrep.2007.04.002} {\bibfield  {journal}
  {\bibinfo  {journal} {Phys. Rep.}\ }\textbf {\bibinfo {volume} {447}},\
  \bibinfo {pages} {1} (\bibinfo {year} {2007})}\BibitemShut {NoStop}%
\bibitem [{\citenamefont {Hayot}\ and\ \citenamefont
  {Jayaprakash}(1997{\natexlab{a}})}]{HayotStrFn}%
  \BibitemOpen
  \bibfield  {author} {\bibinfo {author} {\bibfnamefont {F.}~\bibnamefont
  {Hayot}}\ and\ \bibinfo {author} {\bibfnamefont {C.}~\bibnamefont
  {Jayaprakash}},\ }\href {\doibase 10.1103/PhysRevE.56.227} {\bibfield
  {journal} {\bibinfo  {journal} {Phys. Rev. E}\ }\textbf {\bibinfo {volume}
  {56}},\ \bibinfo {pages} {227} (\bibinfo {year}
  {1997}{\natexlab{a}})}\BibitemShut {NoStop}%
\bibitem [{\citenamefont {Hayot}\ and\ \citenamefont
  {Jayaprakash}(1996)}]{HayotMultifrac}%
  \BibitemOpen
  \bibfield  {author} {\bibinfo {author} {\bibfnamefont {F.}~\bibnamefont
  {Hayot}}\ and\ \bibinfo {author} {\bibfnamefont {C.}~\bibnamefont
  {Jayaprakash}},\ }\href {\doibase 10.1103/PhysRevE.54.4681} {\bibfield
  {journal} {\bibinfo  {journal} {Phys. Rev. E}\ }\textbf {\bibinfo {volume}
  {54}},\ \bibinfo {pages} {4681} (\bibinfo {year} {1996})}\BibitemShut
  {NoStop}%
\bibitem [{\citenamefont {Mitra}\ \emph {et~al.}(2005)\citenamefont {Mitra},
  \citenamefont {Bec}, \citenamefont {Pandit},\ and\ \citenamefont
  {Frisch}}]{BurgMitra}%
  \BibitemOpen
  \bibfield  {author} {\bibinfo {author} {\bibfnamefont {D.}~\bibnamefont
  {Mitra}}, \bibinfo {author} {\bibfnamefont {J.}~\bibnamefont {Bec}}, \bibinfo
  {author} {\bibfnamefont {R.}~\bibnamefont {Pandit}}, \ and\ \bibinfo {author}
  {\bibfnamefont {U.}~\bibnamefont {Frisch}},\ }\href {\doibase
  10.1103/PhysRevLett.94.194501} {\bibfield  {journal} {\bibinfo  {journal}
  {Phys. Rev. Lett.}\ }\textbf {\bibinfo {volume} {94}},\ \bibinfo {pages}
  {194501} (\bibinfo {year} {2005})}\BibitemShut {NoStop}%
\bibitem [{\citenamefont {Kolmogorov}(1941)}]{K41}%
  \BibitemOpen
  \bibfield  {author} {\bibinfo {author} {\bibfnamefont {A.~N.}\ \bibnamefont
  {Kolmogorov}},\ }\href@noop {} {\bibfield  {journal} {\bibinfo  {journal}
  {Dokl. Akad. Nauk. SSSR}\ }\textbf {\bibinfo {volume} {31}},\ \bibinfo
  {pages} {5385} (\bibinfo {year} {1941})}\BibitemShut {NoStop}%
\bibitem [{\citenamefont {Biferale}\ \emph {et~al.}(2011)\citenamefont
  {Biferale}, \citenamefont {Calzavarini},\ and\ \citenamefont
  {Toschi}}]{BiferaleCorr}%
  \BibitemOpen
  \bibfield  {author} {\bibinfo {author} {\bibfnamefont {L.}~\bibnamefont
  {Biferale}}, \bibinfo {author} {\bibfnamefont {E.}~\bibnamefont
  {Calzavarini}}, \ and\ \bibinfo {author} {\bibfnamefont {F.}~\bibnamefont
  {Toschi}},\ }\href {\doibase 10.1063/1.3623466} {\bibfield  {journal}
  {\bibinfo  {journal} {Phys. Fluids}\ }\textbf {\bibinfo {volume} {23}},\
  \bibinfo {pages} {085107} (\bibinfo {year} {2011})}\BibitemShut {NoStop}%
\bibitem [{\citenamefont {Mitra}\ and\ \citenamefont
  {Pandit}(2005)}]{mitra2005dynamics}%
  \BibitemOpen
  \bibfield  {author} {\bibinfo {author} {\bibfnamefont {D.}~\bibnamefont
  {Mitra}}\ and\ \bibinfo {author} {\bibfnamefont {R.}~\bibnamefont {Pandit}},\
  }\href@noop {} {\bibfield  {journal} {\bibinfo  {journal} {Physical review
  letters}\ }\textbf {\bibinfo {volume} {95}},\ \bibinfo {pages} {144501}
  (\bibinfo {year} {2005})}\BibitemShut {NoStop}%
\bibitem [{\citenamefont {Kraichnan}(1968)}]{Kraichnan}%
  \BibitemOpen
  \bibfield  {author} {\bibinfo {author} {\bibfnamefont {R.~H.}\ \bibnamefont
  {Kraichnan}},\ }\href {\doibase 10.1063/1.1692063} {\bibfield  {journal}
  {\bibinfo  {journal} {Phys. Fluids}\ }\textbf {\bibinfo {volume} {11}},\
  \bibinfo {pages} {945} (\bibinfo {year} {1968})}\BibitemShut {NoStop}%
\bibitem [{\citenamefont {Falkovich}\ \emph {et~al.}(2001)\citenamefont
  {Falkovich}, \citenamefont {Gaw\c{e}dzki},\ and\ \citenamefont
  {Vergassola}}]{falkovich2001particles}%
  \BibitemOpen
  \bibfield  {author} {\bibinfo {author} {\bibfnamefont {G.}~\bibnamefont
  {Falkovich}}, \bibinfo {author} {\bibfnamefont {K.}~\bibnamefont
  {Gaw\c{e}dzki}}, \ and\ \bibinfo {author} {\bibfnamefont {M.}~\bibnamefont
  {Vergassola}},\ }\href@noop {} {\bibfield  {journal} {\bibinfo  {journal}
  {Reviews of modern Physics}\ }\textbf {\bibinfo {volume} {73}},\ \bibinfo
  {pages} {913} (\bibinfo {year} {2001})}\BibitemShut {NoStop}%
\bibitem [{\citenamefont {Biferale}(2003)}]{biferale2003shell}%
  \BibitemOpen
  \bibfield  {author} {\bibinfo {author} {\bibfnamefont {L.}~\bibnamefont
  {Biferale}},\ }\href@noop {} {\bibfield  {journal} {\bibinfo  {journal}
  {Annual review of fluid mechanics}\ }\textbf {\bibinfo {volume} {35}},\
  \bibinfo {pages} {441} (\bibinfo {year} {2003})}\BibitemShut {NoStop}%
\bibitem [{\citenamefont {Chen}\ and\ \citenamefont {Kraichnan}(1989)}]{Sweep}%
  \BibitemOpen
  \bibfield  {author} {\bibinfo {author} {\bibfnamefont {S.}~\bibnamefont
  {Chen}}\ and\ \bibinfo {author} {\bibfnamefont {R.~H.}\ \bibnamefont
  {Kraichnan}},\ }\href {\doibase 10.1063/1.857475} {\bibfield  {journal}
  {\bibinfo  {journal} {Phys. Fluids A: Fluid Dyn.}\ }\textbf {\bibinfo
  {volume} {1}},\ \bibinfo {pages} {2019} (\bibinfo {year} {1989})}\BibitemShut
  {NoStop}%
\bibitem [{\citenamefont {Belinicher}\ and\ \citenamefont {L'vov}(1987)}]{QL1}%
  \BibitemOpen
  \bibfield  {author} {\bibinfo {author} {\bibfnamefont {V.~I.}\ \bibnamefont
  {Belinicher}}\ and\ \bibinfo {author} {\bibfnamefont {V.~S.}\ \bibnamefont
  {L'vov}},\ }\href@noop {} {\bibfield  {journal} {\bibinfo  {journal} {Sov.
  Phys. JETP}\ }\textbf {\bibinfo {volume} {66}},\ \bibinfo {pages} {303}
  (\bibinfo {year} {1987})}\BibitemShut {NoStop}%
\bibitem [{\citenamefont {Bateman}(1915)}]{bateman1915some}%
  \BibitemOpen
  \bibfield  {author} {\bibinfo {author} {\bibfnamefont {H.}~\bibnamefont
  {Bateman}},\ }\href@noop {} {\bibfield  {journal} {\bibinfo  {journal}
  {Monthly Weather Review}\ }\textbf {\bibinfo {volume} {43}},\ \bibinfo
  {pages} {163} (\bibinfo {year} {1915})}\BibitemShut {NoStop}%
\bibitem [{\citenamefont {Burgers}(1974)}]{burgers2013nonlinear}%
  \BibitemOpen
  \bibfield  {author} {\bibinfo {author} {\bibfnamefont {J.~M.}\ \bibnamefont
  {Burgers}},\ }\href@noop {} {\emph {\bibinfo {title} {The nonlinear diffusion
  equation: asymptotic solutions and statistical problems}}}\ (\bibinfo
  {publisher} {Springer Netherlands},\ \bibinfo {year} {1974})\BibitemShut
  {NoStop}%
\bibitem [{\citenamefont {Hopf}(1950)}]{hopf1950partial}%
  \BibitemOpen
  \bibfield  {author} {\bibinfo {author} {\bibfnamefont {E.}~\bibnamefont
  {Hopf}},\ }\href@noop {} {\emph {\bibinfo {title} {The Partial Differential
  Equation $u_t+uu_x=\mu u_{xx}$}}},\ \bibinfo {type} {Tech. Rep.}\ (\bibinfo
  {institution} {Indiana University at Bloomington},\ \bibinfo {year}
  {1950})\BibitemShut {NoStop}%
\bibitem [{\citenamefont {Cole}(1951)}]{cole1951quasi}%
  \BibitemOpen
  \bibfield  {author} {\bibinfo {author} {\bibfnamefont {J.~D.}\ \bibnamefont
  {Cole}},\ }\href@noop {} {\bibfield  {journal} {\bibinfo  {journal}
  {Quarterly of applied mathematics}\ }\textbf {\bibinfo {volume} {9}},\
  \bibinfo {pages} {225} (\bibinfo {year} {1951})}\BibitemShut {NoStop}%
\bibitem [{\citenamefont {Gurbatov}\ \emph {et~al.}(1997)\citenamefont
  {Gurbatov}, \citenamefont {Simdyankin}, \citenamefont {Aurell}, \citenamefont
  {Frisch},\ and\ \citenamefont {T\'{o}th}}]{gurbatov1997}%
  \BibitemOpen
  \bibfield  {author} {\bibinfo {author} {\bibfnamefont {S.~N.}\ \bibnamefont
  {Gurbatov}}, \bibinfo {author} {\bibfnamefont {S.~I.}\ \bibnamefont
  {Simdyankin}}, \bibinfo {author} {\bibfnamefont {E.}~\bibnamefont {Aurell}},
  \bibinfo {author} {\bibfnamefont {U.}~\bibnamefont {Frisch}}, \ and\ \bibinfo
  {author} {\bibfnamefont {G.}~\bibnamefont {T\'{o}th}},\ }\href {\doibase
  10.1017/S0022112097006241} {\bibfield  {journal} {\bibinfo  {journal} {J.
  Fluid Mech.}\ }\textbf {\bibinfo {volume} {344}},\ \bibinfo {pages} {339}
  (\bibinfo {year} {1997})}\BibitemShut {NoStop}%
\bibitem [{\citenamefont {Vergassola}\ \emph {et~al.}(1994)\citenamefont
  {Vergassola}, \citenamefont {Dubrulle}, \citenamefont {Frisch},\ and\
  \citenamefont {Noullez}}]{Vergassola}%
  \BibitemOpen
  \bibfield  {author} {\bibinfo {author} {\bibfnamefont {M.}~\bibnamefont
  {Vergassola}}, \bibinfo {author} {\bibfnamefont {B.}~\bibnamefont
  {Dubrulle}}, \bibinfo {author} {\bibfnamefont {U.}~\bibnamefont {Frisch}}, \
  and\ \bibinfo {author} {\bibfnamefont {A.}~\bibnamefont {Noullez}},\ }\href
  {https://ui.adsabs.harvard.edu/abs/1994A&A...289..325V} {\bibfield  {journal}
  {\bibinfo  {journal} {Astron. Astrophys.}\ }\textbf {\bibinfo {volume}
  {289}},\ \bibinfo {pages} {325} (\bibinfo {year} {1994})}\BibitemShut
  {NoStop}%
\bibitem [{\citenamefont {Gurbatov}\ and\ \citenamefont
  {Saichev}(1984)}]{Adhesion}%
  \BibitemOpen
  \bibfield  {author} {\bibinfo {author} {\bibfnamefont {S.~N.}\ \bibnamefont
  {Gurbatov}}\ and\ \bibinfo {author} {\bibfnamefont {A.~I.}\ \bibnamefont
  {Saichev}},\ }\href {\doibase https://doi.org/10.1007/BF01036611} {\bibfield
  {journal} {\bibinfo  {journal} {Radiophys. Quantum Electron.}\ }\textbf
  {\bibinfo {volume} {27}},\ \bibinfo {pages} {303} (\bibinfo {year}
  {1984})}\BibitemShut {NoStop}%
\bibitem [{\citenamefont {Hayot}\ and\ \citenamefont
  {Jayaprakash}(1997{\natexlab{b}})}]{HayotScaling}%
  \BibitemOpen
  \bibfield  {author} {\bibinfo {author} {\bibfnamefont {F.}~\bibnamefont
  {Hayot}}\ and\ \bibinfo {author} {\bibfnamefont {C.}~\bibnamefont
  {Jayaprakash}},\ }\href {\doibase 10.1103/PhysRevE.56.4259} {\bibfield
  {journal} {\bibinfo  {journal} {Phys. Rev. E}\ }\textbf {\bibinfo {volume}
  {56}},\ \bibinfo {pages} {4259} (\bibinfo {year}
  {1997}{\natexlab{b}})}\BibitemShut {NoStop}%
\bibitem [{\citenamefont {Chekhlov}\ and\ \citenamefont
  {Yakhot}(1995)}]{Chekhlov}%
  \BibitemOpen
  \bibfield  {author} {\bibinfo {author} {\bibfnamefont {A.}~\bibnamefont
  {Chekhlov}}\ and\ \bibinfo {author} {\bibfnamefont {V.}~\bibnamefont
  {Yakhot}},\ }\href {\doibase 10.1103/PhysRevE.51.R2739} {\bibfield  {journal}
  {\bibinfo  {journal} {Phys. Rev. E}\ }\textbf {\bibinfo {volume} {51}},\
  \bibinfo {pages} {R2739} (\bibinfo {year} {1995})}\BibitemShut {NoStop}%
\bibitem [{\citenamefont {Kardar}\ \emph {et~al.}(1986)\citenamefont {Kardar},
  \citenamefont {Parisi},\ and\ \citenamefont {Zhang}}]{KPZmain}%
  \BibitemOpen
  \bibfield  {author} {\bibinfo {author} {\bibfnamefont {M.}~\bibnamefont
  {Kardar}}, \bibinfo {author} {\bibfnamefont {G.}~\bibnamefont {Parisi}}, \
  and\ \bibinfo {author} {\bibfnamefont {Y.-C.}\ \bibnamefont {Zhang}},\ }\href
  {\doibase 10.1103/PhysRevLett.56.889} {\bibfield  {journal} {\bibinfo
  {journal} {Phys. Rev. Lett.}\ }\textbf {\bibinfo {volume} {56}},\ \bibinfo
  {pages} {889} (\bibinfo {year} {1986})}\BibitemShut {NoStop}%
\bibitem [{\citenamefont {Barab\'{a}si}\ and\ \citenamefont
  {Stanley}(1995)}]{KPZ1}%
  \BibitemOpen
  \bibfield  {author} {\bibinfo {author} {\bibfnamefont {A.~L.}\ \bibnamefont
  {Barab\'{a}si}}\ and\ \bibinfo {author} {\bibfnamefont {H.~E.}\ \bibnamefont
  {Stanley}},\ }\href
  {https://books.google.co.in/books?hl=en&lr=&id=W4SqcNr8PLYC&oi=fnd&pg=PR15&ots=eIPPBOz4ny&sig=SVtW1xOXa4GbCef3PNDp1mnhU8w&redir_esc=y#v=onepage&q&f=false}
  {\emph {\bibinfo {title} {Fractal Concepts in Surface Growth}}}\ (\bibinfo
  {publisher} {Cambridge University Press},\ \bibinfo {year}
  {1995})\BibitemShut {NoStop}%
\bibitem [{\citenamefont {Polyakov}(1995)}]{PolyakovRG}%
  \BibitemOpen
  \bibfield  {author} {\bibinfo {author} {\bibfnamefont {A.~M.}\ \bibnamefont
  {Polyakov}},\ }\href {\doibase 10.1103/PhysRevE.52.6183} {\bibfield
  {journal} {\bibinfo  {journal} {Phys. Rev. E}\ }\textbf {\bibinfo {volume}
  {52}},\ \bibinfo {pages} {6183} (\bibinfo {year} {1995})}\BibitemShut
  {NoStop}%
\bibitem [{\citenamefont {Canuto}\ \emph {et~al.}(2007)\citenamefont {Canuto},
  \citenamefont {Hussaini}, \citenamefont {Quarteroni},\ and\ \citenamefont
  {Zang}}]{canuto2007spectral}%
  \BibitemOpen
  \bibfield  {author} {\bibinfo {author} {\bibfnamefont {C.}~\bibnamefont
  {Canuto}}, \bibinfo {author} {\bibfnamefont {M.~Y.}\ \bibnamefont
  {Hussaini}}, \bibinfo {author} {\bibfnamefont {A.}~\bibnamefont
  {Quarteroni}}, \ and\ \bibinfo {author} {\bibfnamefont {T.~A.}\ \bibnamefont
  {Zang}},\ }\href@noop {} {\emph {\bibinfo {title} {Spectral methods:
  fundamentals in single domains}}}\ (\bibinfo  {publisher} {Springer Science
  \& Business Media},\ \bibinfo {year} {2007})\BibitemShut {NoStop}%
\bibitem [{\citenamefont {Kloeden}\ and\ \citenamefont
  {Platen}(1992)}]{kloeden1992stochastic}%
  \BibitemOpen
  \bibfield  {author} {\bibinfo {author} {\bibfnamefont {P.~E.}\ \bibnamefont
  {Kloeden}}\ and\ \bibinfo {author} {\bibfnamefont {E.}~\bibnamefont
  {Platen}},\ }in\ \href@noop {} {\emph {\bibinfo {booktitle} {Numerical
  Solution of Stochastic Differential Equations}}}\ (\bibinfo  {publisher}
  {Springer},\ \bibinfo {year} {1992})\ pp.\ \bibinfo {pages}
  {103--160}\BibitemShut {NoStop}%
\bibitem [{\citenamefont {Higham}(2001)}]{higham2001algorithmic}%
  \BibitemOpen
  \bibfield  {author} {\bibinfo {author} {\bibfnamefont {D.~J.}\ \bibnamefont
  {Higham}},\ }\href@noop {} {\bibfield  {journal} {\bibinfo  {journal} {SIAM
  review}\ }\textbf {\bibinfo {volume} {43}},\ \bibinfo {pages} {525} (\bibinfo
  {year} {2001})}\BibitemShut {NoStop}%
\bibitem [{Note1()}]{Note1}%
  \BibitemOpen
  \bibinfo {note} {This holds unless the shocks are distributed on a fractal
  set, which is not the case here.}\BibitemShut {Stop}%
\bibitem [{\citenamefont {Boldyrev}\ \emph {et~al.}(2004)\citenamefont
  {Boldyrev}, \citenamefont {Linde},\ and\ \citenamefont {Polyakov}}]{pdfU}%
  \BibitemOpen
  \bibfield  {author} {\bibinfo {author} {\bibfnamefont {S.}~\bibnamefont
  {Boldyrev}}, \bibinfo {author} {\bibfnamefont {T.}~\bibnamefont {Linde}}, \
  and\ \bibinfo {author} {\bibfnamefont {A.}~\bibnamefont {Polyakov}},\ }\href
  {\doibase 10.1103/PhysRevLett.93.184503} {\bibfield  {journal} {\bibinfo
  {journal} {Phys. Rev. Lett.}\ }\textbf {\bibinfo {volume} {93}},\ \bibinfo
  {pages} {184503} (\bibinfo {year} {2004})}\BibitemShut {NoStop}%
\bibitem [{\citenamefont {Kaneda}\ \emph {et~al.}(1999)\citenamefont {Kaneda},
  \citenamefont {Ishihara},\ and\ \citenamefont {Gotoh}}]{IshiharaDyn}%
  \BibitemOpen
  \bibfield  {author} {\bibinfo {author} {\bibfnamefont {Y.}~\bibnamefont
  {Kaneda}}, \bibinfo {author} {\bibfnamefont {T.}~\bibnamefont {Ishihara}}, \
  and\ \bibinfo {author} {\bibfnamefont {K.}~\bibnamefont {Gotoh}},\ }\href
  {\doibase 10.1063/1.870077} {\bibfield  {journal} {\bibinfo  {journal} {Phys.
  Fluids}\ }\textbf {\bibinfo {volume} {11}},\ \bibinfo {pages} {2154}
  (\bibinfo {year} {1999})}\BibitemShut {NoStop}%
\bibitem [{\citenamefont {Bogaevsky}(2002)}]{Perestroika1}%
  \BibitemOpen
  \bibfield  {author} {\bibinfo {author} {\bibfnamefont {I.~A.}\ \bibnamefont
  {Bogaevsky}},\ }\href {\doibase
  https://doi.org/10.1016/S0167-2789(02)00652-8} {\bibfield  {journal}
  {\bibinfo  {journal} {Physica D: Nonlin. Phenomena}\ }\textbf {\bibinfo
  {volume} {173}},\ \bibinfo {pages} {1} (\bibinfo {year} {2002})}\BibitemShut
  {NoStop}%
\bibitem [{\citenamefont {Ni}\ \emph {et~al.}(2013)\citenamefont {Ni},
  \citenamefont {Shi},\ and\ \citenamefont {Chen}}]{1Dcomp}%
  \BibitemOpen
  \bibfield  {author} {\bibinfo {author} {\bibfnamefont {Q.}~\bibnamefont
  {Ni}}, \bibinfo {author} {\bibfnamefont {Y.}~\bibnamefont {Shi}}, \ and\
  \bibinfo {author} {\bibfnamefont {S.}~\bibnamefont {Chen}},\ }\href {\doibase
  10.1063/1.4816294} {\bibfield  {journal} {\bibinfo  {journal} {Phys. Fluids}\
  }\textbf {\bibinfo {volume} {25}},\ \bibinfo {pages} {075106} (\bibinfo
  {year} {2013})}\BibitemShut {NoStop}%
\bibitem [{\citenamefont {Lele}(1994)}]{Compress1}%
  \BibitemOpen
  \bibfield  {author} {\bibinfo {author} {\bibfnamefont {S.~K.}\ \bibnamefont
  {Lele}},\ }\href {\doibase 10.1146/annurev.fl.26.010194.001235} {\bibfield
  {journal} {\bibinfo  {journal} {Ann. Rev. Fluid Mech.}\ }\textbf {\bibinfo
  {volume} {26}},\ \bibinfo {pages} {211} (\bibinfo {year} {1994})}\BibitemShut
  {NoStop}%
\bibitem [{\citenamefont {Canuto}(1997)}]{Compress2}%
  \BibitemOpen
  \bibfield  {author} {\bibinfo {author} {\bibfnamefont {V.~M.}\ \bibnamefont
  {Canuto}},\ }\href {\doibase 10.1086/304175} {\bibfield  {journal} {\bibinfo
  {journal} {Astrophys. J.}\ }\textbf {\bibinfo {volume} {482}},\ \bibinfo
  {pages} {827} (\bibinfo {year} {1997})}\BibitemShut {NoStop}%
\bibitem [{\citenamefont {Kritsuk}\ \emph {et~al.}(2013)\citenamefont
  {Kritsuk}, \citenamefont {Wagner},\ and\ \citenamefont {Norman}}]{Compress3}%
  \BibitemOpen
  \bibfield  {author} {\bibinfo {author} {\bibfnamefont {A.~G.}\ \bibnamefont
  {Kritsuk}}, \bibinfo {author} {\bibfnamefont {R.}~\bibnamefont {Wagner}}, \
  and\ \bibinfo {author} {\bibfnamefont {M.~L.}\ \bibnamefont {Norman}},\
  }\href {\doibase 10.1017/jfm.2013.342} {\bibfield  {journal} {\bibinfo
  {journal} {J. Fluid Mech.}\ }\textbf {\bibinfo {volume} {729}},\ \bibinfo
  {pages} {R1} (\bibinfo {year} {2013})}\BibitemShut {NoStop}%
\bibitem [{\citenamefont {Sekundov}\ and\ \citenamefont
  {Yakubovskii}(2019)}]{Compress4}%
  \BibitemOpen
  \bibfield  {author} {\bibinfo {author} {\bibfnamefont {A.~N.}\ \bibnamefont
  {Sekundov}}\ and\ \bibinfo {author} {\bibfnamefont {K.~Y.}\ \bibnamefont
  {Yakubovskii}},\ }\href {\doibase 10.1134/S0015462819020113} {\bibfield
  {journal} {\bibinfo  {journal} {Fluid Dyn.}\ }\textbf {\bibinfo {volume}
  {54}},\ \bibinfo {pages} {184} (\bibinfo {year} {2019})}\BibitemShut
  {NoStop}%
\bibitem [{\citenamefont {Wang}\ \emph {et~al.}(2012)\citenamefont {Wang},
  \citenamefont {Shi}, \citenamefont {Wang}, \citenamefont {Xiao},
  \citenamefont {He},\ and\ \citenamefont {Chen}}]{Compress5}%
  \BibitemOpen
  \bibfield  {author} {\bibinfo {author} {\bibfnamefont {J.}~\bibnamefont
  {Wang}}, \bibinfo {author} {\bibfnamefont {Y.}~\bibnamefont {Shi}}, \bibinfo
  {author} {\bibfnamefont {L.-P.}\ \bibnamefont {Wang}}, \bibinfo {author}
  {\bibfnamefont {Z.}~\bibnamefont {Xiao}}, \bibinfo {author} {\bibfnamefont
  {X.~T.}\ \bibnamefont {He}}, \ and\ \bibinfo {author} {\bibfnamefont
  {S.}~\bibnamefont {Chen}},\ }\href {\doibase 10.1017/jfm.2012.474} {\bibfield
   {journal} {\bibinfo  {journal} {J. Fluid Mech.}\ }\textbf {\bibinfo {volume}
  {713}},\ \bibinfo {pages} {588} (\bibinfo {year} {2012})}\BibitemShut
  {NoStop}%
\bibitem [{\citenamefont {Wang}\ \emph {et~al.}(2017)\citenamefont {Wang},
  \citenamefont {Gotoh},\ and\ \citenamefont {Watanabe}}]{Compress6}%
  \BibitemOpen
  \bibfield  {author} {\bibinfo {author} {\bibfnamefont {J.}~\bibnamefont
  {Wang}}, \bibinfo {author} {\bibfnamefont {T.}~\bibnamefont {Gotoh}}, \ and\
  \bibinfo {author} {\bibfnamefont {T.}~\bibnamefont {Watanabe}},\ }\href
  {\doibase 10.1103/PhysRevFluids.2.053401} {\bibfield  {journal} {\bibinfo
  {journal} {Phys. Rev. Fluids}\ }\textbf {\bibinfo {volume} {2}},\ \bibinfo
  {pages} {053401} (\bibinfo {year} {2017})}\BibitemShut {NoStop}%
\bibitem [{\citenamefont {Yang}\ \emph {et~al.}(2016)\citenamefont {Yang},
  \citenamefont {Wang}, \citenamefont {Shi}, \citenamefont {Xiao},
  \citenamefont {He},\ and\ \citenamefont {Chen}}]{CompressLag1}%
  \BibitemOpen
  \bibfield  {author} {\bibinfo {author} {\bibfnamefont {Y.}~\bibnamefont
  {Yang}}, \bibinfo {author} {\bibfnamefont {J.}~\bibnamefont {Wang}}, \bibinfo
  {author} {\bibfnamefont {Y.}~\bibnamefont {Shi}}, \bibinfo {author}
  {\bibfnamefont {Z.}~\bibnamefont {Xiao}}, \bibinfo {author} {\bibfnamefont
  {X.~T.}\ \bibnamefont {He}}, \ and\ \bibinfo {author} {\bibfnamefont
  {S.}~\bibnamefont {Chen}},\ }\href {\doibase 10.1017/jfm.2015.681} {\bibfield
   {journal} {\bibinfo  {journal} {J. Fluid Mech.}\ }\textbf {\bibinfo {volume}
  {786}},\ \bibinfo {pages} {R6} (\bibinfo {year} {2016})}\BibitemShut
  {NoStop}%
\bibitem [{\citenamefont {Konstandin}\ \emph {et~al.}(2012)\citenamefont
  {Konstandin}, \citenamefont {Federrath}, \citenamefont {Klessen},\ and\
  \citenamefont {Schmidt}}]{CompressLag2}%
  \BibitemOpen
  \bibfield  {author} {\bibinfo {author} {\bibfnamefont {L.}~\bibnamefont
  {Konstandin}}, \bibinfo {author} {\bibfnamefont {C.}~\bibnamefont
  {Federrath}}, \bibinfo {author} {\bibfnamefont {R.~S.}\ \bibnamefont
  {Klessen}}, \ and\ \bibinfo {author} {\bibfnamefont {W.}~\bibnamefont
  {Schmidt}},\ }\href {\doibase 10.1017/jfm.2011.503} {\bibfield  {journal}
  {\bibinfo  {journal} {J. Fluid Mech.}\ }\textbf {\bibinfo {volume} {692}},\
  \bibinfo {pages} {183} (\bibinfo {year} {2012})}\BibitemShut {NoStop}%
\bibitem [{\citenamefont {Zhang}\ \emph {et~al.}(2016)\citenamefont {Zhang},
  \citenamefont {Liu}, \citenamefont {Ma},\ and\ \citenamefont
  {Xiao}}]{CompressPl1}%
  \BibitemOpen
  \bibfield  {author} {\bibinfo {author} {\bibfnamefont {Q.}~\bibnamefont
  {Zhang}}, \bibinfo {author} {\bibfnamefont {H.}~\bibnamefont {Liu}}, \bibinfo
  {author} {\bibfnamefont {Z.}~\bibnamefont {Ma}}, \ and\ \bibinfo {author}
  {\bibfnamefont {Z.}~\bibnamefont {Xiao}},\ }\href {\doibase
  10.1063/1.4948810} {\bibfield  {journal} {\bibinfo  {journal} {Phys. Fluids}\
  }\textbf {\bibinfo {volume} {28}},\ \bibinfo {pages} {055104} (\bibinfo
  {year} {2016})}\BibitemShut {NoStop}%
\bibitem [{\citenamefont {Xia}\ \emph {et~al.}(2016)\citenamefont {Xia},
  \citenamefont {Shi}, \citenamefont {Zhang},\ and\ \citenamefont
  {Chen}}]{CompressPl2}%
  \BibitemOpen
  \bibfield  {author} {\bibinfo {author} {\bibfnamefont {Z.}~\bibnamefont
  {Xia}}, \bibinfo {author} {\bibfnamefont {Y.}~\bibnamefont {Shi}}, \bibinfo
  {author} {\bibfnamefont {Q.}~\bibnamefont {Zhang}}, \ and\ \bibinfo {author}
  {\bibfnamefont {S.}~\bibnamefont {Chen}},\ }\href {\doibase
  10.1063/1.4939794} {\bibfield  {journal} {\bibinfo  {journal} {Phys. Fluids}\
  }\textbf {\bibinfo {volume} {28}},\ \bibinfo {pages} {016103} (\bibinfo
  {year} {2016})}\BibitemShut {NoStop}%
\bibitem [{\citenamefont {Boldyrev}\ \emph
  {et~al.}(2002{\natexlab{a}})\citenamefont {Boldyrev}, \citenamefont
  {Nordlund},\ and\ \citenamefont {Padoan}}]{CompressAstro1}%
  \BibitemOpen
  \bibfield  {author} {\bibinfo {author} {\bibfnamefont {S.}~\bibnamefont
  {Boldyrev}}, \bibinfo {author} {\bibfnamefont {A.}~\bibnamefont {Nordlund}},
  \ and\ \bibinfo {author} {\bibfnamefont {P.}~\bibnamefont {Padoan}},\ }\href
  {\doibase 10.1103/PhysRevLett.89.031102} {\bibfield  {journal} {\bibinfo
  {journal} {Phys. Rev. Lett.}\ }\textbf {\bibinfo {volume} {89}},\ \bibinfo
  {pages} {031102} (\bibinfo {year} {2002}{\natexlab{a}})}\BibitemShut
  {NoStop}%
\bibitem [{\citenamefont {Boldyrev}\ \emph
  {et~al.}(2002{\natexlab{b}})\citenamefont {Boldyrev}, \citenamefont
  {Nordlund},\ and\ \citenamefont {Padoan}}]{CompressAstro2}%
  \BibitemOpen
  \bibfield  {author} {\bibinfo {author} {\bibfnamefont {S.}~\bibnamefont
  {Boldyrev}}, \bibinfo {author} {\bibfnamefont {A.}~\bibnamefont {Nordlund}},
  \ and\ \bibinfo {author} {\bibfnamefont {P.}~\bibnamefont {Padoan}},\ }\href
  {\doibase 10.1086/340758} {\bibfield  {journal} {\bibinfo  {journal}
  {Astrophys. J.}\ }\textbf {\bibinfo {volume} {573}},\ \bibinfo {pages} {678}
  (\bibinfo {year} {2002}{\natexlab{b}})}\BibitemShut {NoStop}%
\bibitem [{\citenamefont {Bruno}\ and\ \citenamefont
  {Carbone}(2013)}]{CompressAstro3}%
  \BibitemOpen
  \bibfield  {author} {\bibinfo {author} {\bibfnamefont {R.}~\bibnamefont
  {Bruno}}\ and\ \bibinfo {author} {\bibfnamefont {V.}~\bibnamefont
  {Carbone}},\ }\href {\doibase 0.12942/lrsp-2013-2} {\bibfield  {journal}
  {\bibinfo  {journal} {Livin. Rev. Sol. Phys.}\ }\textbf {\bibinfo {volume}
  {10}} (\bibinfo {year} {2013}),\ 0.12942/lrsp-2013-2}\BibitemShut {NoStop}%
\bibitem [{\citenamefont {Kr{\"u}ger}\ \emph
  {et~al.}(2017{\natexlab{a}})\citenamefont {Kr{\"u}ger}, \citenamefont
  {Haugen}, \citenamefont {Mitra},\ and\ \citenamefont
  {L{\o}v{\aa}s}}]{kruger2017effect}%
  \BibitemOpen
  \bibfield  {author} {\bibinfo {author} {\bibfnamefont {J.}~\bibnamefont
  {Kr{\"u}ger}}, \bibinfo {author} {\bibfnamefont {N.~E.}\ \bibnamefont
  {Haugen}}, \bibinfo {author} {\bibfnamefont {D.}~\bibnamefont {Mitra}}, \
  and\ \bibinfo {author} {\bibfnamefont {T.}~\bibnamefont {L{\o}v{\aa}s}},\
  }\href@noop {} {\bibfield  {journal} {\bibinfo  {journal} {Proceedings of the
  Combustion Institute}\ }\textbf {\bibinfo {volume} {36}},\ \bibinfo {pages}
  {2333} (\bibinfo {year} {2017}{\natexlab{a}})}\BibitemShut {NoStop}%
\bibitem [{\citenamefont {Kr{\"u}ger}\ \emph
  {et~al.}(2017{\natexlab{b}})\citenamefont {Kr{\"u}ger}, \citenamefont
  {Haugen},\ and\ \citenamefont {L{\o}v{\aa}s}}]{kruger2017correlation}%
  \BibitemOpen
  \bibfield  {author} {\bibinfo {author} {\bibfnamefont {J.}~\bibnamefont
  {Kr{\"u}ger}}, \bibinfo {author} {\bibfnamefont {N.~E.~L.}\ \bibnamefont
  {Haugen}}, \ and\ \bibinfo {author} {\bibfnamefont {T.}~\bibnamefont
  {L{\o}v{\aa}s}},\ }\href@noop {} {\bibfield  {journal} {\bibinfo  {journal}
  {Combustion and Flame}\ }\textbf {\bibinfo {volume} {185}},\ \bibinfo {pages}
  {160} (\bibinfo {year} {2017}{\natexlab{b}})}\BibitemShut {NoStop}%
\bibitem [{\citenamefont {Drossel}\ and\ \citenamefont {Kardar}(2002)}]{PRW1}%
  \BibitemOpen
  \bibfield  {author} {\bibinfo {author} {\bibfnamefont {B.}~\bibnamefont
  {Drossel}}\ and\ \bibinfo {author} {\bibfnamefont {M.}~\bibnamefont
  {Kardar}},\ }\href {\doibase 10.1103/PhysRevB.66.195414} {\bibfield
  {journal} {\bibinfo  {journal} {Phys. Rev. B}\ }\textbf {\bibinfo {volume}
  {66}},\ \bibinfo {pages} {195414} (\bibinfo {year} {2002})}\BibitemShut
  {NoStop}%
\bibitem [{\citenamefont {Chin}(2002)}]{PRW2}%
  \BibitemOpen
  \bibfield  {author} {\bibinfo {author} {\bibfnamefont {C.-S.}\ \bibnamefont
  {Chin}},\ }\href {\doibase 10.1103/PhysRevE.66.021104} {\bibfield  {journal}
  {\bibinfo  {journal} {Phys. Rev. E}\ }\textbf {\bibinfo {volume} {66}},\
  \bibinfo {pages} {021104} (\bibinfo {year} {2002})}\BibitemShut {NoStop}%
\bibitem [{\citenamefont {Nagar}\ \emph {et~al.}(2006)\citenamefont {Nagar},
  \citenamefont {Majumdar},\ and\ \citenamefont {Barma}}]{PRW3}%
  \BibitemOpen
  \bibfield  {author} {\bibinfo {author} {\bibfnamefont {A.}~\bibnamefont
  {Nagar}}, \bibinfo {author} {\bibfnamefont {S.~N.}\ \bibnamefont {Majumdar}},
  \ and\ \bibinfo {author} {\bibfnamefont {M.}~\bibnamefont {Barma}},\ }\href
  {\doibase 10.1103/PhysRevE.74.021124} {\bibfield  {journal} {\bibinfo
  {journal} {Phys. Rev. E}\ }\textbf {\bibinfo {volume} {74}},\ \bibinfo
  {pages} {021124} (\bibinfo {year} {2006})}\BibitemShut {NoStop}%
\bibitem [{\citenamefont {Singha}\ and\ \citenamefont {Barma}(2018)}]{PRW4}%
  \BibitemOpen
  \bibfield  {author} {\bibinfo {author} {\bibfnamefont {T.}~\bibnamefont
  {Singha}}\ and\ \bibinfo {author} {\bibfnamefont {M.}~\bibnamefont {Barma}},\
  }\href {\doibase 10.1103/PhysRevE.98.052148} {\bibfield  {journal} {\bibinfo
  {journal} {Phys. Rev. E}\ }\textbf {\bibinfo {volume} {98}},\ \bibinfo
  {pages} {052148} (\bibinfo {year} {2018})}\BibitemShut {NoStop}%
\end{thebibliography}%

\pagebreak
\onecolumngrid
\appendix
\section{Numerical methods}\label{app:A}
\begin{table*}[h]
\centering{}
\begin{tabular}{c c c c c c c c c c c c}
\hline
$\rm{Run}$ & $N$ & $\nu$ & $\delta t$ & $\Li$ & $\urms$ & Re & $\eta$ & $T_L$ & $\tau_\eta$ & $\kmax\eta$ & $N_p$\\
\hline
\hline
$\rm{R1}$& $2^{16}$ & $10^{-6}$ & $5\times10^{-5}$ & $0.87$ & $0.10$ & $8.7\times10^4$ & $1.44\times10^{-4}$ & $8.70$ & $2.1\times10^{-2}$ & $3.14$ & $2^{16}$\\
$\rm{R2}$& $2^{20}$ & $10^{-7}$ & $2\times10^{-6}$ & $0.85$ & $0.22$ & $1.6\times10^6$ & $3.09\times10^{-5}$ & $3.86$ & $9.5\times10^{-3}$ & $10.8$ & $2^{20}$\\
\hline
\end{tabular}\caption{\label{tab:parameters}{\small{}Parameters for our direct numerical simulations (DNSs): $N$ is the number of collocation points, $\nu$ the kinematic viscosity, $\delta t$ the time step, $\Li\equiv\left(\sum_{k}\left|\hat{u}(k)\right|^{2}/k\right)/\left(\sum_{k}\left|\hat{u}(k)\right|^{2}\right)$ the integral length scale, $\urms\equiv\sum_{k}\left|\hat{u}(k)\right|^{2}$ the root-mean-square velocity, ${\rm Re}\equiv \urms\Li/\nu$  the integral-scale Reynolds number, $\eta\equiv\left(\frac{\nu^3}{\epsilon}\right)^{1/4}$ the dissipation length scale, where $\epsilon$ is the energy dissipation rate, $k_{max}$ the dealiasing cutoff, $T_L\equiv \Li/\urms$ the large eddy turnover time, $\tau_\eta\equiv\left(\frac{\nu}{\epsilon}\right)^{1/2}$ the dissipation time scale, and $N_p$ the number of tracers used in our Lagrangian studies.}}
\end{table*}
\begin{figure*}[h]
\begin{centering}
\includegraphics[scale=0.26]{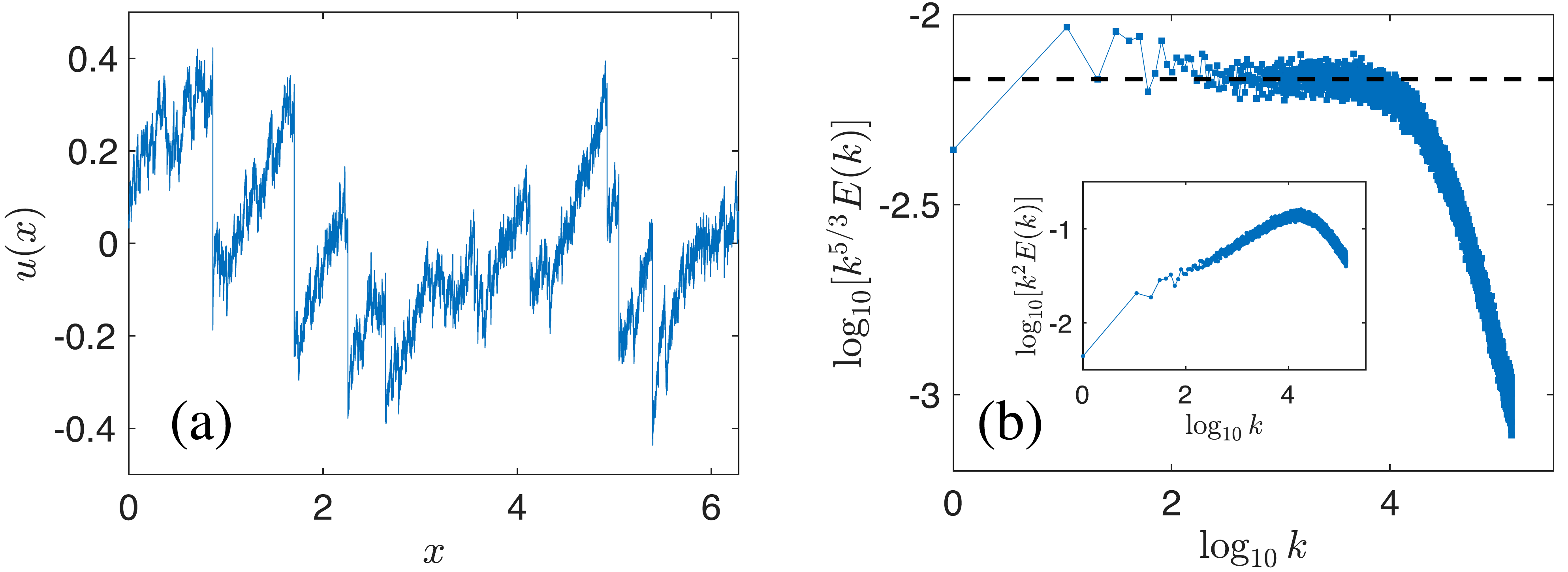}
\par\end{centering}
\caption{\label{fig:staticquantities}{\small{}(a) A plot versus $x$ of the velocity profile $u(x)$ at a representative time in the non-equilibrium statistically steady state (NESS) of Eqs. \eqref{eq:Burgers} and \eqref{eq:noise}. (b) Log-log plot of the compensated energy spectrum, $k^{5/3}E(k)$ in this NESS, plotted up until the forcing-cutoff wave number $k_c=N/8$; the horizontal dashed line denotes the K41 scaling, $E(k)\sim k^{-5/3}$. \textit{Inset:} Energy-dissipation spectrum $k^2E(k)$ plotted on a log-log scale up until $k_c=N/8$; the prominent peak in the graph indicates that the dissipation range is well-resolved. Both plots are obtained from run R2.}}
\end{figure*}

A representative velocity profile in the NESS and the compensated energy spectrum are shown in Fig.~\ref{fig:staticquantities}. The parameters for our DNSs are listed in Table~\ref{tab:parameters}.\\
\paragraph{\textbf{Quasi-Lagrangian (QL) velocity $V(x,t)$:}}

The QL velocities are calculated in the reference frame of a tracer which is advected by the local mean flow. In order to calculate the QL velocities, we load a single Lagrangian particle (tracer) at $x_0=\pi$, i.e., in the middle of the simulation domain. At every iteration, we calculate the displacement $R(t)$ of the tracer, initially at the position $x_0$, by using the following equations:
\begin{eqnarray}
\frac{dR(t)}{dt}&=&u(x_R,t) \, ; \nonumber \\ x_R(t)&=&x_0+R(t) \,;
\end{eqnarray}
$x_R(t)$ is the position of the tracer at time $t$ and $u(x_R,t)$ is the Eulerian flow velocity at the location of the tracer. We solve the equation of motion for $R(t)$ by using the forward-Euler method. We then obtain the QL velocity 
\begin{equation}
V(x,t)=u(x+R(t),t)\,.
\end{equation}
For off-grid positions of a tracer, we need to find its velocity via interpolation. To circumvent interpolation errors, we calculate the components of the QL velocities exactly in Fourier space by using the relation $\hat{V}(k,t)=\hat{u}(k,\,t)e^{ikR(t)}$. After this we transform the QL velocities back to real space, such that $V(x,t)=\sum_k \hat{V}(k,t)e^{ikx}=\sum_k \hat{u}(k,t)e^{ik[x+R(t)]}=u(x+R(t),t)$.

\section{Collapse-time exponents: Dependence on different representative initial configurations in the NESS}\label{app:B}
\begin{figure}[H]
    \centering
    \includegraphics[scale=0.55]{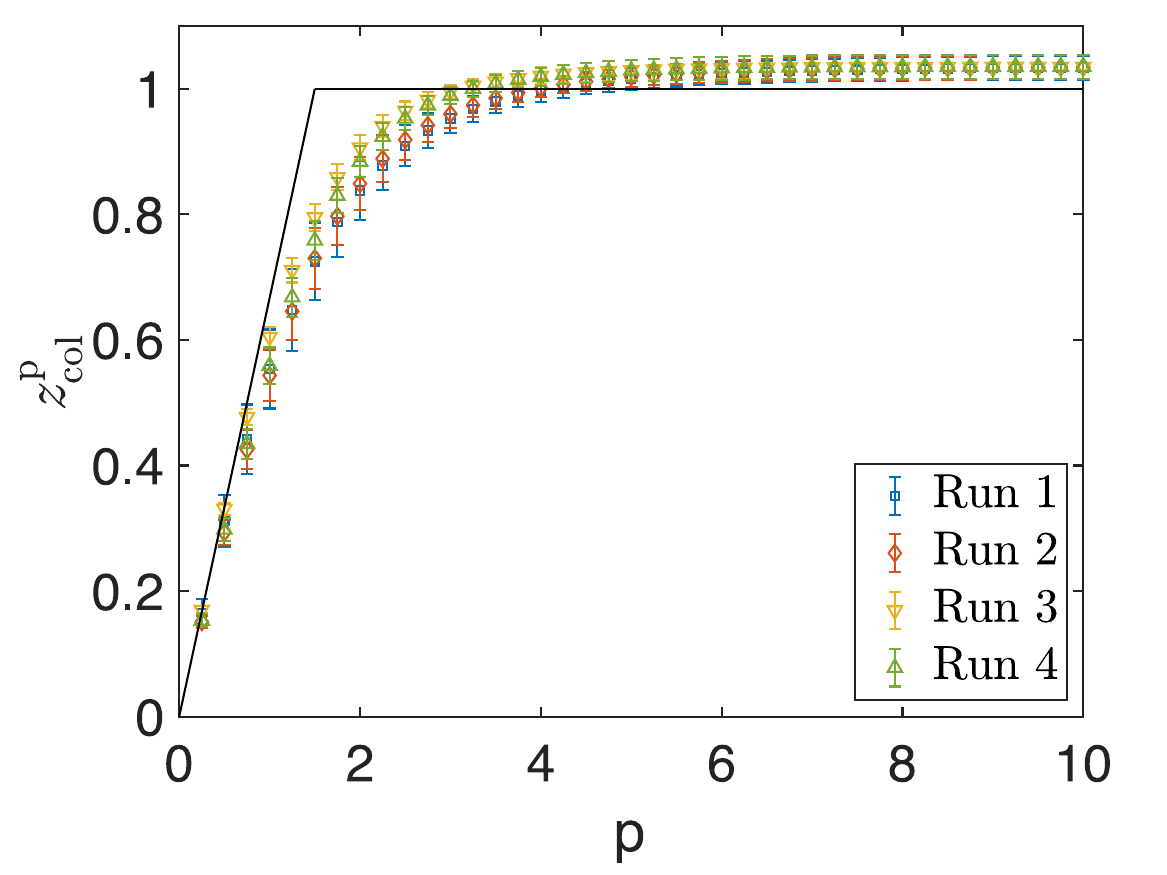}
    \caption{\small Plots of $\zpfm$ versus $p$ for runs starting from different initial NESS's at a resolution of $N=2^{16}$ collocation points; the black lines denote the theoretical bifractal prediction [Eq. \eqref{eq:zpfm2}]. The exponents from these different runs agree with each other, within error bars.}
    \label{fig:manyfm}
\end{figure}

We have also checked whether the interval-collapse-time exponents, $\zpfm$, depend on using different representative initial conditions, from the NESS of Eqs. \eqref{eq:Burgers} and \eqref{eq:noise}, into which we  introduce $N_p$ equi-spaced tracers to start the runs with particles. We find that, given a large-enough spatial resolution, this dependence is small, and it lies within the errors bars that we have given for $\zpfm$. We show this explicitly in Fig.~\ref{fig:manyfm} for four representative DNSs with $N=2^{16}$ collocation points.

\section{PDF $\mathcal{P}(v_0)$ of $v_0$} \label{app:C}
\begin{figure}[H]
    \centering
    \includegraphics[scale=0.55]{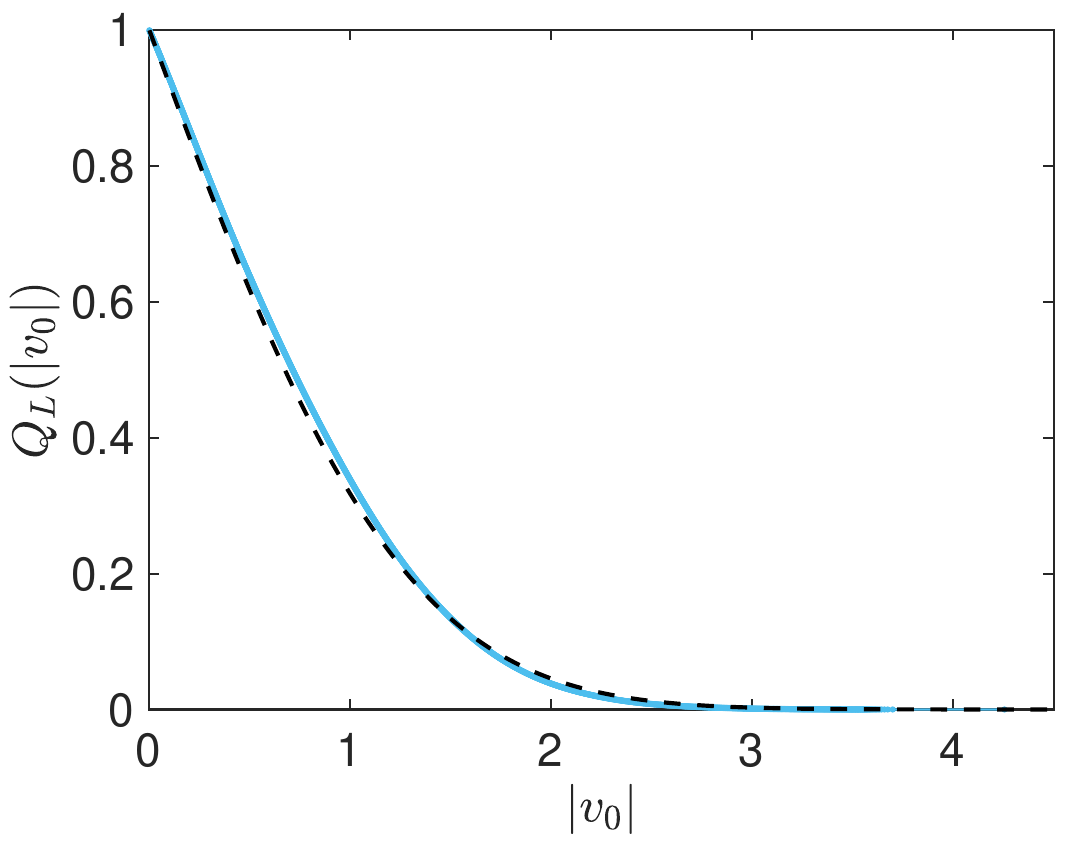}
    \caption{\small Plot of the normalized c-CPDF $Q_L(|v_0|)$ of the magnitude of $v_0$ versus $|v_0|$, represented in blue; the black dashed curve is the c-CPDF of a standard normal distribution. The two curves match closely with each other, thereby justifying our assumption in the main text.}
    \label{fig:u0pdf}
\end{figure}

We assume that $v_0$ follows a standard normal distribution, i.e., $\mathcal{P}(v_0)\sim e^{-v_0^2/2}$, as in incompressible fluid turbulence. In order to justify our assumption, we calculate the complementary CPDF (c-CPDF), $Q_L(|v_0|)$, of the magnitude of $v_0$ by using the rank-order method. This circumvents binning errors which can affect the numerical determination of the PDF. In Fig. \ref{fig:u0pdf}, we observe that the graph of $Q_L(|v_0|)$ is in good agreement with that of the c-CPDF of the standard normal distribution, thereby justifying our assumption. A similar result has been obtained earlier in Ref.~\cite{pdfU}.

\section{Scaling of the moments of $\tfm$ with $r$}\label{app:D}

The order-$p$ moment of $\tau$ is 
\begin{equation}
    \Tpfm(r)=\left<\tau^p\right>=\int_0^\infty \tau^p\Phi(\tau)d\tau \,.
    \label{eq:Eqn14}
\end{equation}
$\Tpfm$ has contributions from three different terms on the RHS of Eq. \eqref{eq:PDFtcol}, so we analyze it  term by term.

\textit{(A)} The contribution to $\Tpfm$ from the first term on the RHS of Eq. \eqref{eq:PDFtcol} is
\begin{equation}
    \TpfmA(r) \sim r^{2/3}\int_0^\infty\tau^{p-2}\exp\left[-\frac{r^{4/3}}{2\tau^2}\right]d\tau \,.
    \label{eq:Eqn18}
\end{equation}
By making the substitution $s=\sqrt{2}\tau/r^{2/3}$, we get
\begin{equation}
    \TpfmA(r) \sim r^{2p/3}\int_0^\infty s^{p-2}\exp\left[-\frac{1}{s^2}\right]ds \,,
    \label{eq:Eqn19}
\end{equation}
which implies,
\begin{equation}
    \TpfmA(r)\sim r^{2p/3} \,.
    \label{eq:Eqn20}
\end{equation}

\textit{(B1)} The contribution from the second term on the RHS of Eq. \eqref{eq:PDFtcol} is
\begin{equation}
    \TpfmBI(r) \sim r^2\int_0^\infty\tau^{p-2}\exp\left[-\frac{r^2}{2\tau^2}\right]d\tau \,.
    \label{eq:Eqn15}
\end{equation}
By making the substitution $s=\sqrt{2}\tau/r$, we get
\begin{equation}
    \TpfmBI(r) \sim r^{p+1}\int_0^{\infty}s^{p-2}\exp\left[-\frac{1}{s^2}\right]ds \,.
    \label{eq:Eqn16}
\end{equation}
This implies that
\begin{equation}
    \TpfmBI(r)\sim r^{p+1} \,.
    \label{eq:Eqn17}
\end{equation}

\textit{(B2)} The contribution from $\Phi_{B2}(\tau)$ is,
\begin{equation}
    \TpfmBII(r) \sim \sum_{\tstar}\left[W_{\tstar}r\int_{\tstar}^\infty\frac{\tau^p}{(\tau-\tstar)^2}\exp\left\{-\frac{r^2}{2(\tau-\tstar)^2}\right\}d\tau\right] \,,
    \label{eq:Eqn21}
\end{equation}
where $W_{\tstar}=w(\tstar)$. By making the substitution $s=r/(\tau-\tstar)$, we get
\begin{equation}
    \TpfmBII(r) \sim \sum_{\tstar}\left[W_{\tstar}\int_0^{\infty}\left(\frac{r}{s}+\tstar\right)^pe^{-s^2/2}ds\right] \,.
    \label{eq:Eqn22}
\end{equation}
We now expand $\left(\frac{r}{s}+\tstar\right)^p$ in a binomial series. In order to ensure its convergence for the different values of $s$, given the values of $\tstar$ and $r$, we write
\begin{equation}
    \left(\tstar+\frac{r}{s}\right)^p=\begin{cases}
    \tstar^p\left(1+\frac{r}{s\tstar}\right)^p &\text{if}\quad s>\frac{r}{\tstar} \,; \\ \\
    \frac{r^p}{s^p}\left(1+\frac{s\tstar}{r}\right)^p &\text{if}\quad s<\frac{r}{\tstar} \,.
    \end{cases}
    \label{eq:Eqn23}
\end{equation}
Then, Eq. \eqref{eq:Eqn22} can be written as
\begin{equation*}
    \TpfmBII(r) \sim \sum_{\tstar}\left[W_{\tstar}\left(I_1+I_2\right)\right] \,,
    \label{eq:Eqn24}
\end{equation*}
where
\begin{equation}
    I_1 = r^p\int_0^{r/\tstar}\frac{1}{s^p}\left(1+\frac{s\tstar}{r}\right)^pe^{-s^2/2}ds = r^p\int_0^{r/\tstar}\left[s^{-p}+\frac{p\tstar}{r}s^{1-p}+p(p-1)\left(\frac{\tstar}{r}\right)^2s^{2-p}+...\right]e^{-s^2/2}ds
    \label{eq:Eqn25}
\end{equation}
and
\begin{equation}
    I_2 = \tstar^p\int_{r/\tstar}^{\infty}\left(1+\frac{r}{s\tstar}\right)^pe^{-s^2/2}ds = \tstar^p\int_{r/\tstar}^{\infty}\left[1+p\frac{r}{s\tstar}+p(p-1)\left(\frac{r}{s\tstar}\right)^2+...\right]e^{-s^2/2}ds \,.
    \label{eq:Eqn26}
\end{equation}

By making the substitution $y=s^2/2$, the integrals over each term in Eq. \eqref{eq:Eqn25} can be be written in terms of the lower incomplete gamma function $\gamma(n,x)=\int_0^xt^{n-1}e^{-t}dt$. We use the asymptotic form of $\gamma(n,x)$, in the limit of $x\to0$, $\gamma(n,x)\sim\frac{x^n}{n}$, because $r\ll1$. We find that the leading-order contribution from $I_1$ scales as $r^2$. 

By making the same substitution in Eq. \eqref{eq:Eqn26}, we find that each one of the integrals in Eq. \eqref{eq:Eqn26} can be written in terms of the upper incomplete gamma function $\Gamma(n,x)=\int_x^{\infty}t^{n-1}e^{-t}dt$. By using the asymptotic properties of the resulting terms in the limit of $r\ll1$, we find that the $r-$dependent leading-order contribution scales linearly with $r$, with logarithmic corrections. As a result,
\begin{equation}
    \TpfmBII(r) \sim C_1r+C_2r\ln r\,,
    \label{eq:Eqn27}
\end{equation}
where $C_1\sim\sum_{\tstar}\left[\frac{p}{2\tstar}(\gamma-2\ln \tstar)\right]$ and $C_2\sim\sum_{\tstar}\left(\frac{p}{\tstar}\right)$ and $\gamma$ the Euler constant. Given the
scaling forms of $\tau$ in Eqs.~\eqref{eq:tauA}, \eqref{eq:tauB1}, and $\eqref{eq:tauB2}$, and $T_L\sim L/v_0$, $\tau\equiv\tfm(\ell)/T_L$, $r\equiv \ell/L$, $\tau\ll 1$ for all collapsing intervals (recall that $\tstar<\tau$), $C_1\gg C_2$ for all $p>0$, whence we have
\begin{equation}
    \TpfmBII(r) \sim r \,.
    \label{eq:Eqn28}
\end{equation}

By combining Eqs. (\ref{eq:Eqn17}), (\ref{eq:Eqn20}) and (\ref{eq:Eqn28}), we obtain:
\begin{equation}
    \Tpfm(r) \sim \begin{cases}
    r^{2p/3} &\text{for}\quad p\leq 3/2 \,;\\
    r &\text{for}\quad p > 3/2 \,.
    \end{cases}
    \label{eq:Eqn29}
\end{equation}
In this way, we are able to extract from $P(\tfm)$ the same scaling behavior of the moments of $\tfm$ as we had obtained from our theoretical arguments given in the main paper.

\newpage
\section{Equal-time structure functions}\label{app:E}
\begin{figure}[h]
    \centering
    \includegraphics[scale=0.25]{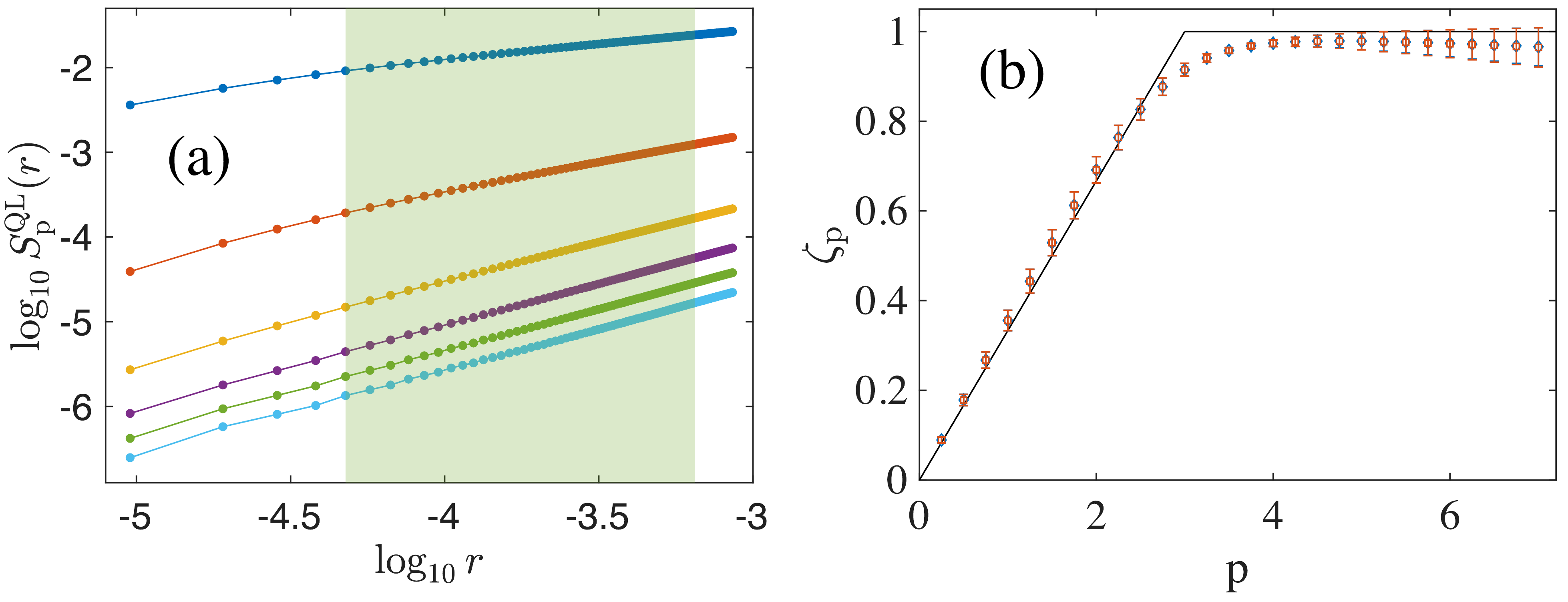}
    \caption{\small Log-log plots of QL equal-time structure functions $\SQLp(r)$ for $p=1$ (dark blue), $p=2$ (red), $p=3$ (yellow), $p=4$ (violet), $p=5$ (green) and $p=6$ (light blue); the green shaded portion indicates the range over which we carry out local-slope analysis for calculating $\zetaQL_{\rm p}$. (b) Equal-time Eulerian (blue diamonds) and QL (red squares) structure function exponents, $\zetaEu_{\rm p}$ and $\zetaQL_{\rm p}$, for different values of $p$; the black lines denote the bifractal exponents given by Eq. \eqref{eq:eqstrfn}c.}
    \label{fig:eqstrs}
\end{figure}

\section{Derivation of the collapse-time exponents from the multifractal model of turbulence}\label{app:F}

In the multifractal model of turbulence, the probability of being within a spatial distance $r$ of a set of fractal dimension $D(h)$ scales as $r^{d-D(h)}$, where $d$ is the embedding dimension of the velocity field. This is the same as the probability, $p_r$, of finding a Lagrangian interval of length $r$ across a set of fractal dimension $D(h)$. By using $\tfm\sim r/\delta u(r)\sim r^{1-h}$, we write
\begin{equation}
    \Tpfm(r)\sim \int_h r^{p(1-h)+1-D(h)}d\mu(h) \sim r^p\int_h r^{-ph+1-D(h)}d\mu(h) \sim r^{\zpfm} \,,
    \label{eq:multifrac}
\end{equation}
where $\mu(h)$ is the weight associated with each $h$ and $h\in[h_{\rm min},h_{\rm max}]$. From the definition of equal-time structure functions $\Sp(r)$ and by using the multifractal model we have
\begin{equation}
    \Sp(r)\sim \int_h r^{ph+1-D(h)}d\mu(h) \sim r^{\zeta_{p}} \,,
    \label{eq:eqstrmultifrac}
\end{equation}
whence we obtain the following bridge relation between $\zpfm$ and $\zeta_{p}$:
\begin{equation}
    \zpfm = p+\zeta_{-p} \,.
    \label{eq:bridgecoll}
\end{equation}
By assuming that $\Sp(r)\equiv \left<\left|\delta u(r)\right|^p\right>$ exists for $p<0$, we have $\zeta_{p}=p/3$ for all $p<0$. Hence, we get the following bridge relation for $\zpfm$:
\begin{equation}
    \zpfm=\frac{2p}{3},\quad{\rm for\;all\;p}.
    \label{eq:zpfm_multifrac}
\end{equation}

This agrees with the one derived earlier from the tracer dynamics [Eq. \eqref{eq:zpfm2}] up to $p=3/2$. As we saw before, the major contribution to $\zpfm$ for $0\leq p \leq 3/2$ mainly comes from those collapsing intervals whose dynamic properties are not significantly affected by shocks. The saturation of $\zpfm$ to unity for $p>3/2$ occurs because of the change in the scaling of the velocity fluctuations across a collapsing interval because of the appearance of a shock at $t\equiv\tstar\ll\tfm$. This cannot be explained, simply, by the multifractal model alone.

\end{document}